 \documentclass{article}
 
\usepackage{authblk} 

\usepackage{amsmath}
\usepackage{graphicx}
\usepackage{txfonts}
\usepackage{natbib}
\usepackage{bm}
\usepackage[addedmarkup=bf]{changes}

\def\beq#1{\begin{equation}\label{#1}}
\def\eeq{\end{equation}}
\def\beqa#1{\begin{eqnarray}\label{#1}}
\def\eeqa{\end{eqnarray}}

\def\Eq#1{Eq.~(\ref{#1})} 
\def\eqn#1{~(\ref{#1})}
\def\myfrac#1#2{\left(\frac{#1}{#2}\right)}

\def\comment#1{\relax}

\def\dfrac#1#2{\displaystyle\frac{\partial #1}{\partial #2}}

\newcommand{\ii}{\mathrm{i}}

\graphicspath{{figures/}}

\title{Magnetorotational instability in Keplerian disks: a non-local approach}
\author[1]{N. Shakura\thanks{E-mail: nikolai.shakura@gmail.com, kpostnov@gmail.com}}
\author[1,2]{K. Postnov}
\author[1,3]{D. Kolesnikov}
\author[1,4]{G. Lipunova}
\affil[1]{\small Moscow State University, Sternberg Astronomical Institute, Universitetskij pr. 13, \newline 119234 Moscow, Russia} %
\affil[2]{\small Kazan Federal University, Kremlyovskaya 18, 420111 Kazan, Russia} %
\affil[3]{\small The Raymond and Beverly Sackler School of Physics and Astronomy,\newline Tel Aviv University, Tel Aviv 69978, Israel}
\affil[4]{\small Max-Planck-Institut f\"ur Radioastronomie, Auf dem H\"ugel 69, 53121 Bonn, Germany}

\begin{document}
\maketitle

\begin{abstract}
We revisit the modal analysis of small perturbations in Keplerian ideal gas flows  leading to magneto-rotational instability (MRI) using the non-local approach. 
We consider the case of constant vertical background magnetic field, as well as the case of radially dependent background Alfv\'en velocity. 
In the case of constant Alfv\'en velocity, MRI modes are described by a Schr\"odinger-like differential equation with some effective potential including 'repulsive' ($1/r^2$) and 'attractive' ($-1/r^3$) terms. 
Taking into account the radial dependence of the background Alfv\'en speed leads to a qualitative change in the shape of the effective potential. 
It is shown that there are no stationary energy levels corresponding to unstable modes $\omega^2 < 0$ in ``shallow'' potentials. 
In thin accretion disks, the wavelength of the disturbance $\lambda=2\piup/k_z$ is smaller than the half-thickness $h$ of the disk only in ``deep'' potentials. 
The limiting value of the background Alfv\'en speed $(c_A)_\mathrm{cr}$, above which the magnetorotational instability does not occur, is found. 
In thin accretion disks with low background Alfv\'en speed $c_A\ll (c_A)_\mathrm{cr}$, the increment of the magnetorotational instability $\omega\approx -\sqrt{3}\mathrm{i}c_Ak_z$ is suppressed compared to the value obtained in the local perturbation analysis.
\end{abstract}

\section*{\small Keywords \textnormal{hydrodynamics, instabilities, magnetic fields}}

\everymath{\displaystyle}
\section{Introduction}
\label{intro}

Shear flows in astrophysical objects, characterized by an inhomogeneous velocity field, are a universal source and agent of energy transport and are closely related to phenomena of turbulence \citep{1973A&A....24..337S,BKL2001,2018book_ch1}, magnetic field generation~\citep{Boneva+2021}, and particle acceleration~\citep{Somov+2003}. 

The stability of shear hydrodynamic flows with respect to small perturbations in a magnetic field in laboratory conditions was first considered papers by E. Velikhov and S. Chandrasekhar~\citep{Velikhov59, 1960PNAS...46..253C}. In the absence of magnetic field, hydrodynamic instability in a rotating shear flow appears when the angular momentum decreases outward from the axis of rotation~\citep{LordRayleigh16}.

 Velikhov and Chandrasekhar showed that in a vertically magnetized, axisymmetric, differentially rotating flow with angular velocity decreasing outward, magnetorotational instability (MRI) is possible.\footnote{ We consider only the case of a vertical background magnetic field. Axisymmetric gas flows with toroidal background magnetic fields in the presence of gravity are subject to Parker instability~\citep{1966ApJ...145..811P}.}

The theory of MRI was applied to astrophysical accretion disks in an influential paper by \cite{1991ApJ...376..214B}, and it is now believed that this instability generates turbulence in accretion disks (see review \cite{1998RvMP...70....1B}). Nonlinear numerical simulations (e.g., \citealt{1995ApJ...440..742H, 2012ApJ...749..189S, 2013ApJ...772..102H}) confirm that MRI can indeed sustain turbulence in accretion disks.

It is believed  that for the study of MRI and analysis of its properties, a local approximation in an ideal incompressible fluid is sufficient, where small perturbations are taken in cylindrical coordinates $r$, $z$, $\varphi$ in the form of plane waves $\propto e^{\,\ii(\omega t-k_r\,r-k_z\,z)}$. 
In this case, the differential MHD equations are transformed into algebraic equations, the dispersion relation for perturbations is found in the form of a biquadratic equation \citep{1991ApJ...376..214B,1998Kato}, and the instability increment does not depend on the magnetic field (see Eqs.~(\ref{e:algebra}) and (\ref{e:omegamax}) below, respectively). 
Taking into account non-ideal effects in this approximation slightly changes the conditions for the occurrence of MRI but leaves qualitatively the same picture ~\citep{2015MNRAS.451.3995S,Zou+2020}.

However, already in the pioneer work of Velikhov~(\citeyear{Velikhov59}), a global analysis of long-wave disturbances in the direction perpendicular to the plane of the main flow was carried out for flows between two  rotating conducting cylinders with a constant angular momentum over radius, $\Omega(r)r^2=const$. Radial perturbations were found from the solution of the Sturm-Liouville boundary value problem (in the VKB approximation). It was shown that such an approach implies a critical magnetic field, above which instability is suppressed, and the dependence on the boundary conditions remains even in the case of extending the outer cylinder to infinity.

Due to the potential importance of MRI in the emergence of turbulence in disk flows (accretion and protoplanetary disks, gaseous disks in galaxies), extensive analytical and numerical investigations of MRI were conducted in the 1990s-2000s in the approximation of incompressible fluid in a homogeneous magnetic field, including a global analysis of this instability. Nonlocal analysis shows that in shear flows, the dispersion equation for the mode $\omega(k)$ contains a term dependent on radius as $\propto  -1/r^2$, which is usually neglected in local modal analysis. As expected, a critical value of magnetic field appears in the global analysis, above which MRI are stabilized~\citep{1992GApFD..66..223P,Kumar-Coleman1994, Gammie-Balbus1994, 1994ApJ...434..206C, 2015MNRAS.453.3257L}. For accretion disks around a central gravitational body, the obtained results depend on the choice of boundary conditions for perturbations at the inner and outer radii of the flow ~\citep{Knobloch1992, 1993A&A...274..667D, 2015MNRAS.453.3257L}. The choice of boundary conditions affects the discretization of local dispersion relations~\citep{2015MNRAS.453.3257L}.

Despite the considerable previous scrutiny of this problem, in this study, we independently and comprehensively perform a non-local linear analysis of MRI in Keplerian accretion disks with an angular velocity $\Omega(r)\sim 1/r^{3/2}$.
We derive a dispersion equation that can be reduced to a Schr\"odinger-like equation with an ``energy'' $E=-k_z^2$ and an effective potential $U(r)$ consisting of two terms: an ``attractive'' term proportional to $-1/r^3$ and a ``repulsive'' term proportional to $1/r^2$. 
In contrast to the local analysis, the effective potential vanishes at a point $r_0$, which depends on the mode frequency $\omega$, the wave number $k_z$, and the background Alfv\'en velocity. 
We examine in detail both cases where the outer radius of the disk, $r_{\rm out}$, is greater or less than $r_0$. 
We numerically solve the boundary value problem for radial boundary conditions corresponding to both rigid and free flow boundaries. 
We emphasize the significance of the position of the flow boundaries relative to the zero points of the effective potential, appearing in the non-local analysis for the Sturm-Liouville boundary problem.
We demonstrate that in ``shallow'' effective potentials, there can be a situation, depending on the position of the inner flow boundary, where MRI do not occur. 
Such a situation is possible for flows around normal stars with large radii. 
Naturally, for flows around compact objects, the potential wells are very deep, and the energy spectrum is nearly continuous. 
We explicitly derive the critical magnetic field value that suppresses MRI  and find the dependence of the MRI increment on the background homogeneous magnetic field. 
We consider for the first time the case of a background magnetic field that varies with radius in a power-law manner. 
We also examine the case of an incompressible fluid with density depending on the radial coordinate, where the equations for small perturbations remain the same as for a constant density, while the effective potential changes.

The structure of the paper is as follows. 
In Section 2, we perform the linear modal analysis for small perturbations in an ideal fluid in the form $\sim f(r)\exp[\ii(\omega t-k_zz)]$. 
In Section~2.5, we derive the algebraic dispersion equation $\omega(k_z)$ and the critical Alfvén velocity below which MRI occurs.  
In Section~3.2, we consider for the first time the MRI in the presence of radially varying vertical magnetic field and demonstrate that in this case, the effective potential can change nontrivially. 
In Section~4, we compare our results with the standard results obtained in the local modal analysis. 
In Appendix~A, we numerically solve the Schrödinger equation for nonlocal perturbations with a constant background Alfvén velocity in the case when the external radius of the flow, $r_{\rm out}$, is greater than the zero radius of the effective potential $r_0$. 
In Appendix~B, we consider the case of $r_{\rm out} < r_0$, when the problem reduces to solving the standard Sturm-Liouville problem with third-type boundary conditions.

\section{Non-local modal analysis}

\subsection{Basic equations}
We consider a differentially rotating ideal fluid in homogeneous vertical magnetic field. Classical results were obtained in papers by E. Velikhov and S. Chandrasekhar who 
studied the stability of sheared hydromagnetic flows \citep{Velikhov59, 1960PNAS...46..253C}.

Equations of motion of ideal MHD fluid read:

1) mass conservation equation
\beq{e:cont}
\frac{\partial \rho}{\partial t}+\nabla(\rho\bm{u})=0\,,
\eeq

2) Euler equation including gravity force and Lorentz force
\beq{e:euler}
\frac{\partial \bm{u}}{\partial t}+(\bm{u}\nabla)\bm{u}=-\frac{1}{\rho}\nabla p -
\nabla \phi_g +
\frac{1}{4\piup\rho}(\nabla\times \bm{B})\times \bm{B}\, 
\eeq
(here $\phi_g$ is the Newtonian gravitational potential\footnote{In principle, one can solve the problem in Schwarzschield metric using the potential $\phi=\frac{c^2}{2}\ln\left(1-\frac{r_g}{r}\right)$, $r_g=2GM/c^2$ see \cite{2018MNRAS.480.4273S}.}), 

3) induction equation
\beq{e:induc}
\frac{\partial \bm{B}}{\partial t}=\nabla\times (\bm{u}\times \bm{B})\,
\eeq

 We will consider adiabatic perturbations with constant entropy
\beq{e:adiab}
\frac{\partial s}{\partial t} + (\bm{u}\nabla) s=0\,.
\eeq
For such adiabatic perturbations, perturbed density variations are zero, $\rho_1=0$, and pressure variations in the energy equation vanish, $p_1=0$ (see, e.g., Appendix in \cite{2015MNRAS.448.3697S}).

We analyze the case of a purely Keplerian rotation where the unperturbed velocity is $v_\phi\equiv u_{\phi,0}=\sqrt{GM/r}$, $u_{r,0}=u_{z,0}=0$. 
We assume that the forces caused by pressure gradient are small and only appear in the perturbed equations.

\subsection{The case of incompressible fluid}

Let us consider small Eulerian perturbations in an ideal incompressible fluid. The velocity components in the background undisturbed flow with velocity $u_{\phi,0}$ will be $u_r, u_\phi, u_z$. The magnetic field can be expressed as $\bm{B}=\bm{B}_0+\bm{b}$, and the pressure as $p_0+p_1$. We will consider the poloidal background field $\bm{B}_0$. We will seek perturbations in the form of $f(r)\exp[\ii (\omega t-k_zz)]$, noting that time $t$ and the coordinate $z$ only appear in the system of equations through the derivative sign.

The choice of perturbations with harmonic functions in the vertical coordinate is dictated by the nature of the problem for disk flows that are confined in the z-coordinate. In such flows (accretion and protoplanetary disks, gas disks in galaxies), the vertical pressure gradient is balanced by the gradient of gravitational force along the $z$-coordinate, distinguishing them from laboratory flows.

The integration of the unperturbed Euler equation over the $z$-coordinate leads to a polytropic density distribution $\rho(r,z)=\rho_{\rm c}(1-(z/z_0)^2)^n$, with a semi-thickness of $z_0={2(n+1)P_{\rm c}}/({\Omega^2(r) \rho_{\rm c}})$, where $\rho_{\rm c}$ and $P_{\rm c}$ are the central density and pressure, and $n$ is the polytropic index ($n=3/2$ for a convectively stable disk). For the polytropic equation of state $n=1/(\gamma-1)$, where $\gamma$ is the adiabatic index, and in the case of an incompressible fluid  $\gamma\to\infty$ and $n = 0$. In this limiting case, the vertical density gradient vanishes, and the model flow represents a Keplerian disk with a constant vertical density limited by the disk's semi-thickness $h$ ($\Pi$-shaped density distribution). The density can vary radially (see Section 3).

For infinitesimal perturbations, the approximation of perturbations with harmonic functions in the $z$-coordinate is suitable for both thin disks ($h/r\ll 1$) and thick disks ($h/r\lesssim 1$), provided that the wavelength of the perturbations is smaller than the disk's half-thickness: $\lambda=2\pi/k_z < h$. Therefore, the final equations for linear perturbations in the case of an incompressible fluid, which will be derived below, are not different from the equations for laboratory plasmas with a constant density along the $z$-coordinate.

For the chosen perturbations, the continuity equation \eqref{e:cont} for an incompressible fluid, $\nabla\bm{u}=0$, in cylindrical coordinates can be expressed as follows:
\beq{iurz}
\frac{1}{r}\frac{\partial}{\partial r}(ru_r)-\ii k_zu_z= \frac{\partial u_r}{\partial r}+\frac{u_r}{r}-\ii k_zu_z=0\, .
\eeq
It should be noted that, in the local approximation,  the small term ${u_r}/{r}$ in the continuity equation is typically neglected. In this case, the perturbations can be sought in the form $\propto \exp[\ii(\omega t - k_rr - k_z z)]$. The equation of magnetic field solenoidality, $\nabla \bm{B} = 0$, can be written in a similar manner:
\beq{bz}
\frac{1}{r}\frac{\partial}{\partial r}(rb_r)-\ii k_zb_z= 0\, .
\eeq

The radial, azimuthal, and vertical components of the Euler equation are, respectively,
\beq{iur}
\ii \omega u_r-2\Omega 
u_\phi=-{\frac{1}{\rho_0}\frac{\partial p_1}{\partial r}}
-\frac{c_A^2}{B_0}\left[\dfrac{b_z}{r}+\ii k_zb_r\right]\,,
\eeq
\beq{iuphi}
\ii \omega u_\phi+\frac{\varkappa^2}{2\Omega}u_r=-\ii \frac{c_A^2}{B_0}k_zb_\phi
\eeq
(here we introduced the epicyclic frequency $
\varkappa^2=4\Omega^2+r\,({{\rm d}\Omega^2}/{{\rm d}r})\equiv ({1}/{r^3})\,{{\rm d}(\Omega^2r^4)}/{{\rm d} r}
$ and unperturbed Alfv\'en velocity $c_A^2=B_0^2/(4\piup\rho_0)$),
\beq{iuz}
\ii \omega u_z=\ii k_z\frac{p_1}{\rho_0}\, .
\eeq

The three components of the induction equation, taking into account the solenoidality of the magnetic field $\nabla \bm{B} = 0$, read:
\beq{ibr}
\ii \omega b_r=-\ii B_0k_zu_r\,,
\eeq
\beq{ibphi} 
\ii \omega b_\phi=-\ii B_0k_zu_\phi+r\frac{{\rm d}\Omega}{{\rm d}r}b_r\,,
\eeq
\beq{ibz}
\mathrm{i}\omega b_z=-B_0\frac{1}{r}\dfrac{ru_r}{r}\,.
\eeq
Let us express all perturbations in terms of the radial perturbations of the magnetic field $b_r$.
Using \Eq{iurz}, \Eq{iuz},  and \Eq{ibr}, we find 
\beq{p1}
\frac{p_1}{\rho_0}=\frac{\omega u_z}{k_z}=\ii \frac{\omega^2}{B_0k_z^3}\,
\frac{1}{r}\,\dfrac{rb_r}{r}\,.
\eeq
From \Eq{ibr} we find
\beq{ur}
u_r=-\frac{\omega}{B_0k_z}b_r\,.
\eeq
Substituting \eqn{ur}, \eqn{ibphi} and \eqn{bz} into \Eq{iuphi} yields:
\beq{uphi}
(c_A^2k_z^2-\omega^2)u_\phi=
\ii \left[\frac{\omega^2\varkappa^2}{2\Omega B_0k_z}-
\frac{c_A^2k_z}{B_0}\,r\,\frac{{\rm d}\Omega}{{\rm d}r}\right]\,b_r\,.
\eeq
Finally, using \eqn{p1}--\eqn{uphi}, from \Eq{iur} we obtain a second-order differential equation for $b_r$:
\beq{e:br}
\left(c_A^2-\frac{\omega^2}{k_z^2}\right)^2\left[
\frac{\partial^2b_r}{\partial r^2}+\frac{\partial }{\partial r}\myfrac{b_r}{r}-b_rk_z^2\right]+
2\Omega\left(\frac{\omega^2}{k_z^2}\frac{\varkappa^2}{2\Omega}-c_A^2r\frac{{\rm d}\Omega}{{\rm d}r}\right)b_r=0\,.
\eeq
This equation should be complemented with boundary conditions, which are determined by the problem's formulation. For instance, E.P. Velikhov considered the flow between two rigid conducting cylinders, where the velocity and magnetic field perturbations were set to zero. In a more realistic scenario applicable to thin accretion disks, boundary conditions need to be chosen at the free boundary. For magnetized disks, various formulations of such boundary conditions are possible (see, for example, the discussion in \citealt{1994ApJ...434..206C}). We will demonstrate that the boundary conditions at the inner and outer edges of a Keplerian thin disk with a magnetic field correspond to the vanishing of Lagrangian variations of pressure. First, let us consider how a homogeneous magnetic field can be created in a thin conducting disk.

\subsection{How to create a uniform magnetic field in a thin disk?}

\begin{figure}
        \centering
        \includegraphics[width=0.9\textwidth]{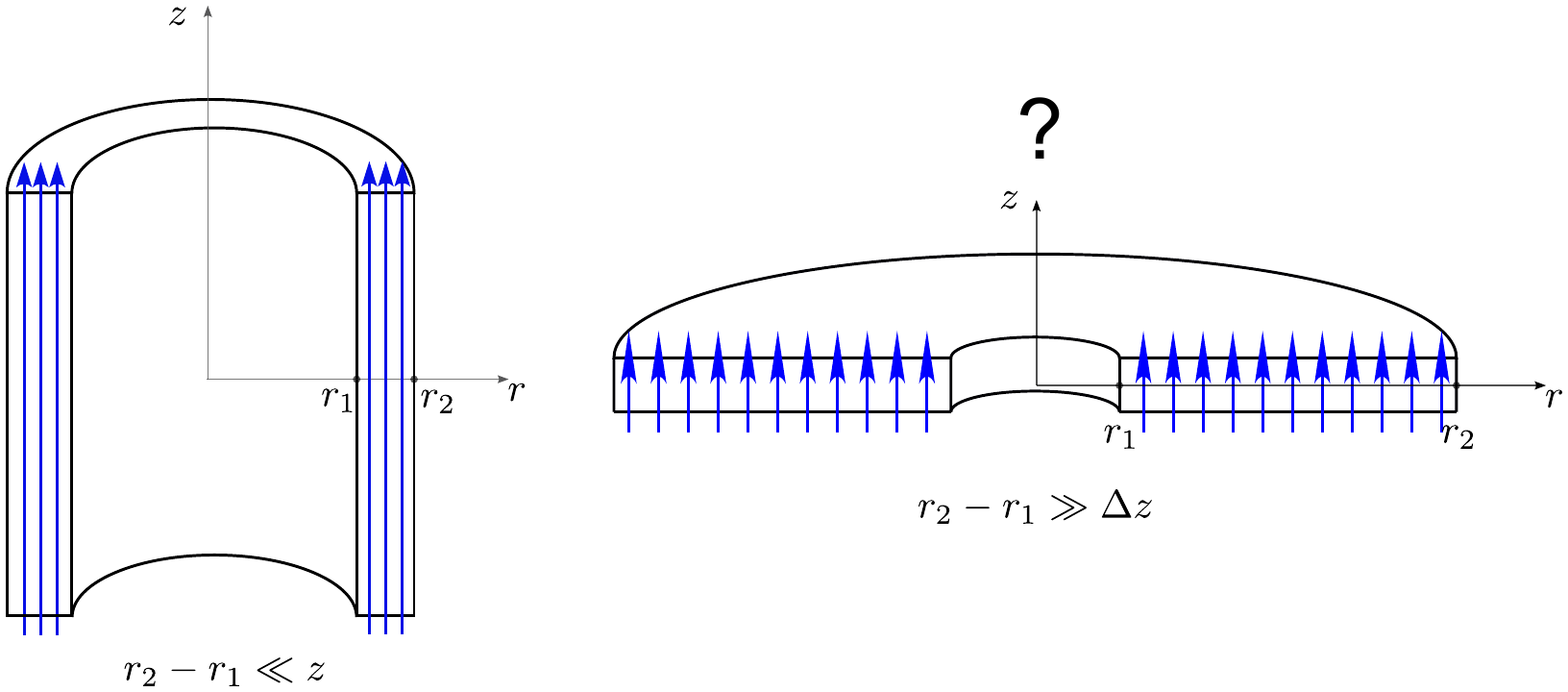}
        \caption{
        (a) Flow in a uniform magnetic field within a narrow gap between two cylinders. (b) A thin extended accretion disk in a uniform vertical magnetic field.}
        \label{f:cylinders}
\end{figure}
    
\begin{figure}
\includegraphics[width=0.49\textwidth]{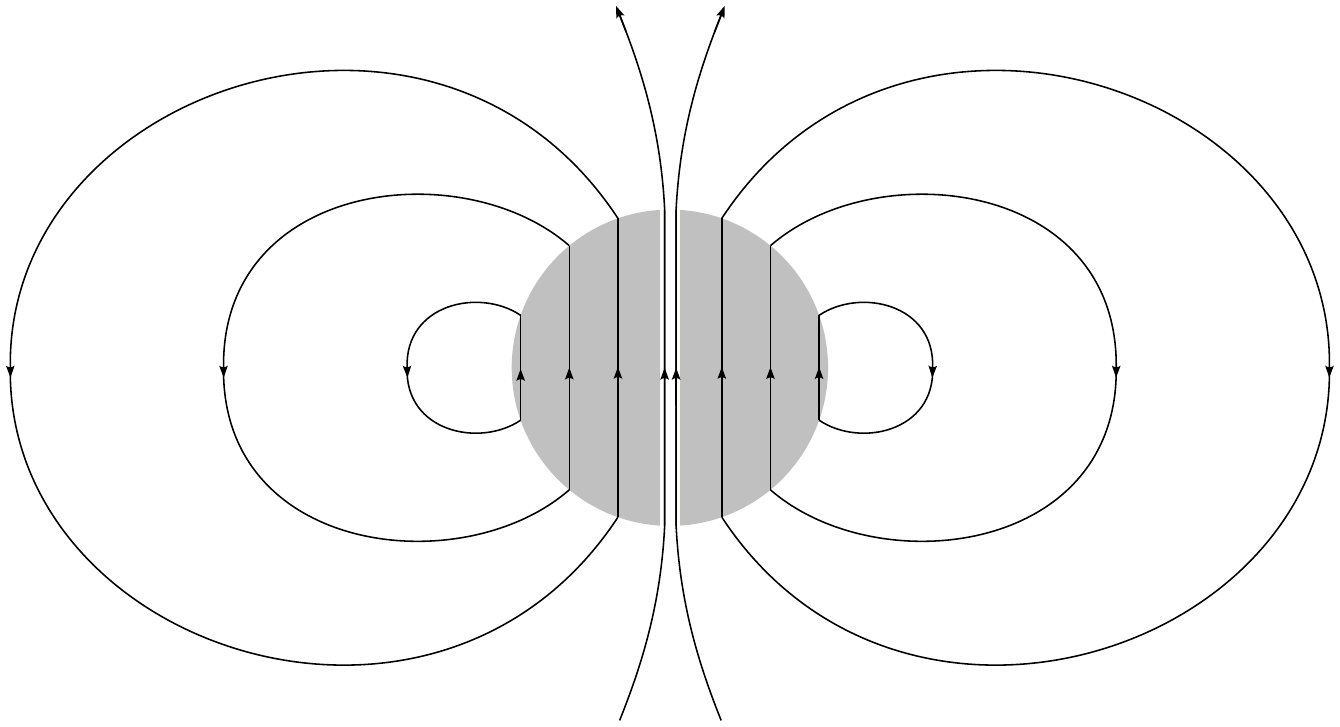}
\includegraphics[width=0.49\textwidth]{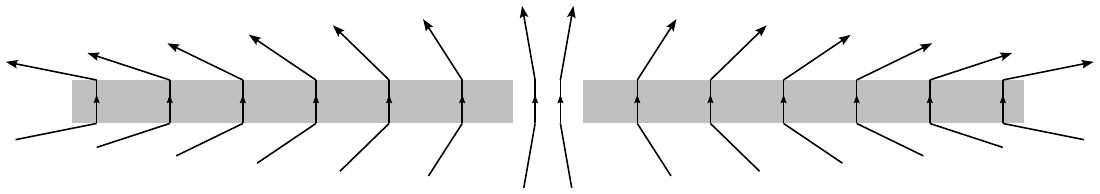}
\caption{
Deformation of a conducting sphere into a disk.The surface Lorentz force $ \bm{j}_{\rm s} \times \bm{B}$ transforms the sphere into an elongated disk in the absence of other forces.
}
\label{f:deform}
\end{figure}
Let us discuss how to create a uniform magnetic field in an accretion disk (see Fig.~\ref{f:cylinders}). First, let us recall that this can be easily achieved in the case of two infinite, well-conducting cylinders (left in Fig.  \ref{f:cylinders}) with currents flowing in opposite directions. However, how can we organize currents in a flat conducting disk to obtain a uniform magnetic field inside it (right in Fig. \ref{f:cylinders})? To answer this question, let us consider the following problem. Suppose we have a well-conducting sphere with surface currents $j_\phi \propto \sin\theta$ such that the field inside the sphere is uniform. We smoothly deform the sphere into a disk while preserving the magnetic flux ${\rm d}\Phi_{\rm s} = B_0\,\cos\theta\, {\rm d}S$, where ${\rm d}S = 2\pi R_0^2\,\sin\theta\, {\rm d}\theta$ is an element of the surface area of the sphere, and $\theta$ is the angle between the surface normal and the magnetic field lines.
In the disk, we have ${\rm d}\Phi_d = B_{\rm d} 2\pi r\,{\rm d}r$ ($B_{\rm d}$ being the field in the disk). Preserving the flux $\Phi_{\rm s} = \Phi_{\rm d}$ implies $2\pi R_0^2\int B_0\,\sin\theta\,\cos\theta\, {\rm d}\theta = 2\pi \int\, B_{\rm d} r \,{\rm d}r$. In the case of a uniform field, we have $R_0^2 B_0\, \sin^2\theta + C = B_{\rm d} r^2$. The constant $C$ must be zero to satisfy the condition $\theta\to 0$, $r\to 0$. For uniform fields and $\theta=\pi/2$, we find the radius of the outer disk: $r_{\rm d} = R_0\sqrt{B_0/B_{\rm d}}$. 
Thus, we have constructed a disk with surface currents that induce a bending of the magnetic field lines on the surface and a uniform magnetic field inside.

Now let us remove a narrow region around the vertical axis from the considered sphere. Since the surface currents vanish near the vertical axis, removing a small region near the axis will not significantly alter the field inside the sphere. Of course, such an operation changes the topology: the sphere transforms into a torus.

Alternatively, the sphere can be deformed into a disk while preserving the magnetic flux, but with a radially variable magnetic field, $B_{\rm d}(r)$. During the deformation process, the changing magnetic field induces an electric field, which in turn strengthens or weakens the surface currents  $\bm{j}_{\rm s}$~(Fig.~\ref{f:deform}). It should be evident that by deforming the sphere while conserving the magnetic flux, any desired dependence of the magnetic field on the radius can be obtained.

\begin{figure}
\begin{center}
\includegraphics[width=0.5\textwidth]{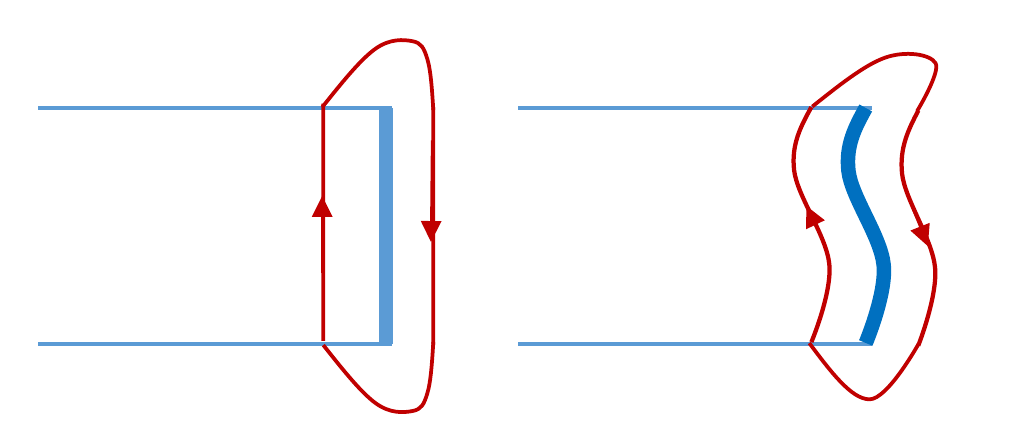}
\caption{
Schematic illustrating the cancellation of the Lorentz force $ \bm{j}_{\rm s} \times \bm{B}$  at the outer boundary of the accretion disk. (a) Unperturbed configuration, with the surface current $\bm{j}_{\rm s}$ flowing perpendicular to the plane of the image. (b) The perturbed case, showing only the projection of the field onto the $r,z$ plane. The perturbed component of the field $b_\phi$  is not shown.
}
\label{f:Lorenz}
\end{center}
\end{figure}

It is worth noting that in the obtained configuration, the surface current at the inner boundary of the disk becomes zero, and consequently, the Lorentz force acting on the surface element, $ \bm{j}_{\rm s} \times \bm{B}$, also vanishes. At the outer boundary, as the magnetic field changes sign when transitioning from the disk to the outside, there is a jump in the magnetic field, and a surface current must flow. The Lorentz force will be determined by the sum of two forces acting on the surface current from the inner and outer magnetic fields. Since, by construction, the outer and inner fields are equal but oppositely directed near the boundary, the total Lorentz force applied to the outer boundary of the disk will be zero (Fig.~\ref{f:Lorenz}).

\subsection{Differential equation for small perturbations}
\label{s:constantCA}

Following  \cite{Velikhov59} and \cite{1960PNAS...46..253C}, it is straightforward to show that in \Eq{e:br} $\omega^2$ must be a real number for a wide range of boundary conditions, i.e., only oscillations ($\omega^2>0$) or exponential growth or decay ($\omega^2<0$) of perturbations are allowed. 
It is convenient to eliminate the first derivative
from \eqn{e:br} using the substitution $\Psi=b_r\sqrt{r}$, in order to obtain a second-order differential equation:
\begin{eqnarray}
   \label{e:shred} 
&
\frac{{\rm d}^2\Psi}{{\rm d}r^2}+\left\{
-k_z^2-\frac{3}{4}\frac{1}{r^2}+
\frac{2\Omega\left[{ \omega^2\,\varkappa^2}/(2\Omega\,k_z^2)-c_A^2r\,({\rm d}\Omega/{\rm d}r)\right]}{\left(c_A^2-{\omega^2}/{k_z^2}\right)^2}
\right\}\times\nonumber\\
& \times\,
\Psi=0\,.
\end{eqnarray}
The boundary conditions for this equation are selected based on physical considerations, ensuring that the Lagrangian pressure perturbations vanish on the free boundary (recall that as shown earlier, the surface Lorentz force is zero on both boundaries):
\begin{equation}\label{e:Lagr}
\Delta p = {\rm \delta} p + (\bm{\xi}\, \nabla)p = 0,
\end{equation}
where $\bm{\xi}$ is the vector of infinitesimal displacement and ${\rm \delta} p = p_1$ is the Eulerian pressure variation. In our problem formulation, Keplerian unperturbed flow is assumed, when the gravitational force is balanced by the centrifugal force, for small displacements $\bm{\xi}$, the second term in the expression for the Lagrangian variation can be neglected (up to terms of order $(h/r)^2$ in a thin accretion disk with semi-thickness $h$), and $\Delta p = p_1$. 
After dividing by the differential of time ${\rm d} t$, the equation \eqref{e:Lagr} can be written as:
\[
\frac{{\rm D}p}{{\rm D}t} = \frac{\partial p}{\partial t} + (\bm{u}\, \nabla) p = 0\, ,
\]
where ${\rm D}/{\rm D}t$ represents the Lagrangian derivative, and $\partial/\partial t$ represents the Eulerian derivative.

At the outer and inner boundaries of the accretion disk with a magnetic field, the boundary conditions can be expressed as $p_1|_{r_\mathrm{in},r_\mathrm{out}}=0$. As implied by Eq.~\eqref{p1}, these boundary conditions lead to homogeneous third-type boundary conditions for the function $\Psi$ at the flow boundaries $r_{\rm in},r_{\rm out}$:

\begin{equation}\label{e:boundary3}
\left.\frac{{\rm d}\Psi}{{\rm d}r}\right|_{r_{\rm in}}+\left.\frac{1}{2}\frac{\Psi}{r}\right|_{r_{\rm in}}=0\, , \quad 
\left.\frac{{\rm d}\Psi}{{\rm d}r}\right|_{r_{\rm out}}+\left.\frac{1}{2}\frac{\Psi}{r}\right|_{r_{\rm out}}=0\,.
\end{equation}
It is worth noting that in the works of E.P. Velikhov (see \citealt{Velikhov59} and \citealt{2015MNRAS.453.3257L}), a boundary value problem with homogeneous first-type conditions at the boundaries ($\Psi|_{r_\mathrm{in}}=0, \,\Psi|_{r_\mathrm{out}}=0$) was considered.

Equation~\eqref{e:shred} can be interpreted as a Schr\"odinger-like equation with "energy" 
\begin{equation}
E=-k_z^2
\end{equation}
and "potential"
\beq{e:U}
U=\frac{3}{4}\frac{1}{r^2}-
\frac{({\omega^2}/{k_z^2})\,\varkappa^2-c_A^2r\,({\rm d}\Omega^2/{{\rm d}r})}{\left(c_A^2-{\omega^2}/{k_z^2}\right)^2}\,.
\eeq
As is well-known (see, for example, \citealt{LL1974,1971pqm..book.....F}), in the region of negative potential values $U$, there exist eigenvalues ("energy levels") with negative energy $E<0$. The existence of levels with negative energy, i.e., solutions for $\bm{b}$, for real $k_z$ and $\omega^2<0$, indicates the instability of the flow. Thus, the form of the potential $U$ determines the region in the flow and the range of values of $\bm{B}$ that allow the development of MRI.

Let us highlight the properties of the potential $U$ that are relevant to our problem. For the Keplerian law with $\Omega^2=GM/r^3$ and $\varkappa^2=\Omega^2$, the potential takes the form:
\beq{e:UK}
U=\frac{3}{4}\frac{1}{r^2}-\frac{GM}{r^3}\,
\frac{{\omega^2}/{k_z^2}+3c_A^2}{\left(c_A^2-{\omega^2}/{k_z^2}\right)^2}\,.
\eeq
\begin{figure}
\begin{center}
\centerline{\includegraphics[width=0.7\textwidth]{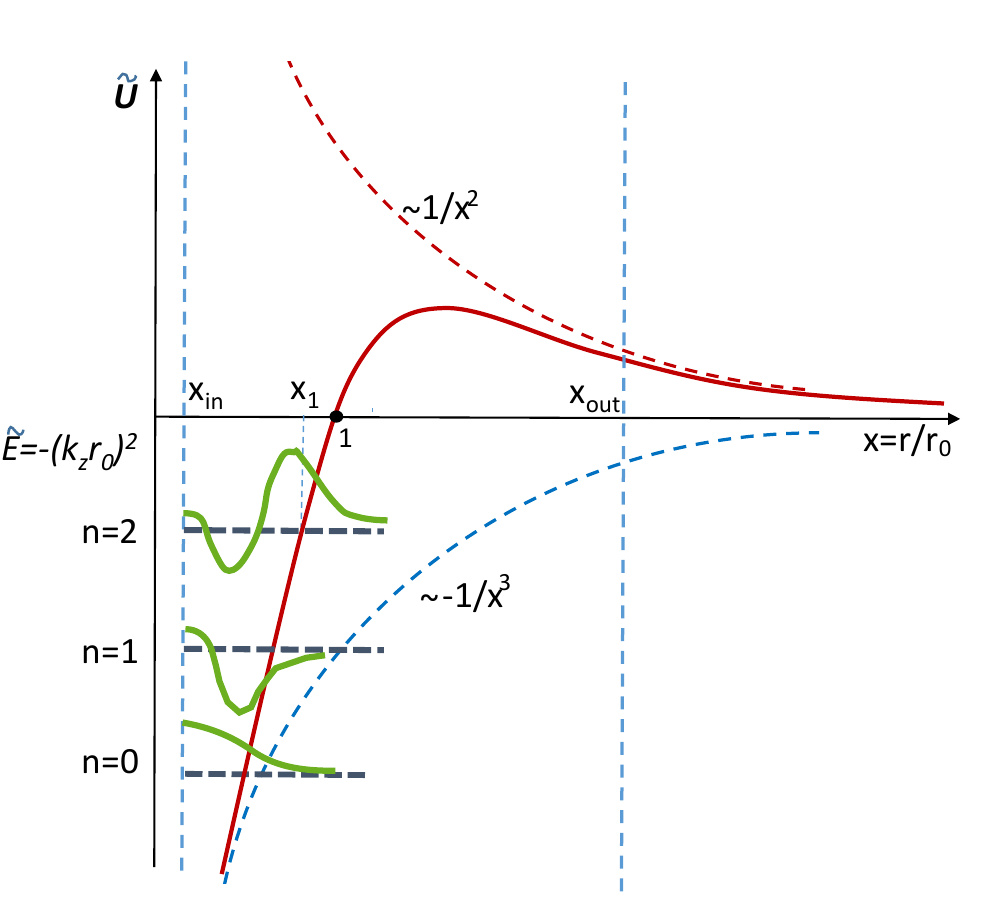}} 
\caption{
Schematic of the effective potential $\tilde U$ with characteristic points $x_\mathrm{in}$ (inner boundary of the flow), $x_1$ (turning point in the potential for a given energy level), $x=1$ (point where the potential becomes zero), and $x_\mathrm{out}$ (outer radius of the flow). The first three energy levels ($n=0,1,2$) and their corresponding eigenfunctions of the problem are shown in green color, in the case where $x_{\rm out}>1$.
}
\label{f:1}
\end{center}
\end{figure}
For unstable modes ($\omega^2<0$), the numerator in the second term of equation \eqref{e:UK} is positive only for perturbations with $\omega^2>-3c_A^2k_z^2$. Otherwise, the potential $U$ is strictly positive, and there are no stable states with negative energy. For $\omega^2<0$, the denominator in equation \eqref{e:UK} does not vanish.
In our problem (see Fig.~\ref{f:1}), there are several characteristic points: the inner radius of the disk $r_\mathrm{in}$, the turning point $r_1$ for a given negative energy $E=-k_z^2$, the zero point of the potential $r_0$, and $r_\mathrm{max}=(3/2) r_0$, where the potential is maximal. It is evident that $r_\mathrm{in}\le r_1\le r_0$.
  
The potential turns to zero at the point 
\beq{e:r0}
r_0=\frac{4}{3}\,G\,M\
\frac{{\omega^2}/{k_z^2}+3c_A^2}{\left(c_A^2-{\omega^2}/{k_z^2}\right)^2}=
\frac{4}{3}\,\frac{G\,M}{c_A^2}\,
\frac{{\omega^2}/{(k_z^2c_A^2)}+3}{\left[1-{\omega^2}/({k_z^2c_A^2})\right]^2}
\,.
\eeq

It is convenient to introduce a dimensionless variable $x=r/r_0$. Then the dimensionless potential can be expressed as follows:
Then the dimensionless potential can be written as
\beq{e:Ux}
\tilde U=U\,r_0^2=\frac{3}{4}\frac{1}{x^2}-\frac{3}{4}\frac{1}{x^3}\,.
\eeq

The dimensionless energy in the Schr\"odinger equation \eqref{e:shred} is equal to
\beq{e:Etilda}
\tilde E=-k_z^2\,r_0^2\,.
\eeq 

The turning point of the potential $U$ is determined by the equation
\beq{e:r1}
-k_z^2-U(r_1)=0\,
\eeq
(see Fig.~\ref{f:1}). 
Beyond the radius $r_1$, the perturbations decay quasi-exponentially. 
In dimensionless units, the turning point $x_1=r_1/r_0$ of the potential $\tilde U$ is determined by the equation:
\beq{e:ntp}
4\tilde E x_1^3-3x_1+3=0, \quad x_1 \le 1\,.
\eeq
For negative energy $\tilde E$, this cubic equation has one real root $x_1=r_1/r_0<1$:
\beq{e:x1cardano}
x_1=\myfrac{3}{-8\tilde E}^{1/3}\left[\left(1-\sqrt{1-\frac{1}{9\tilde E}}\right)^{1/3}+\left(1+\sqrt{1-\frac{1}{9\tilde E}}\right)^{1/3}\right]\,.
\eeq
Numerical solutions of \Eq{e:shred} can be found in Appendix A.
It should be noted that the global non-local analysis of MRI, leading to an equation of the form of a one-dimensional Schr\"odinger Eq.~\eqref{e:shred}, has been investigated by \citet{1993A&A...274..667D,1994ApJ...434..206C} with different boundary conditions.

\subsection{Derivation of the dispersion equation and critical Alfv\'en velocity}
\label{s:de}

Thus, in dimensionless variables, the problem reduces to a Sturm-Liouville problem for the equation
\beq{e:Sh1}
\Psi''-\tilde U\Psi+\tilde E\Psi=0
\eeq
with the potential \eqref{e:Ux} and boundary conditions
\beq{e:boundary3x}
\left.\frac{{\rm d}\Psi}{{\rm d}x}\right|_{x_{\rm in}}+\left.\frac{1}{2}\frac{\Psi}{x}\right|_{x_{\rm in}}=0, \quad 
\left.\frac{{\rm d}\Psi}{{\rm d}x}\right|_{x_{\rm out}}+\left.\frac{1}{2}\frac{\Psi}{x}\right|_{x_{\rm out}}=0\,.
\eeq

Since the potential $\tilde U$ changes sign at $x=1$ ($r=r_0$), when solving this problem, it is necessary to distinguish between two cases: 1) the outer radius of the flow is located beyond the zero point of the potential $x_{\rm out}>1$ ($r_{\rm out}>r_0$), and 2) the flow terminates before reaching the zero point of the potential $x_{\rm out}<1$ ($r_{\rm out}<r_0$) (see details in Appendices A and B, respectively).
From the solution of the boundary value problem, a discrete set of eigenvalues $\tilde E_n$, $n=0,1,2,\ldots$, is obtained,
\beq{e:En}
\tilde E_n=-k_z^2r_0^2\,.
\eeq
In the case of $x_{\rm out}>1$, the number $n$ corresponds to the number of zeros of the eigenfunctions on the interval between the inner boundary and the turning point of the potential for a given level $[x_{\rm in}, x_1(\tilde E_n)]$.

From Eq.~\eqref{e:En}, for each level with negative energy $\tilde E_n$, we obtain a dispersion equation
\beq{e:r0disp}
r_0=\frac{\sqrt{-\tilde E_n}}{k_z}=\frac{4}{3}\,
\frac{GM}{c_A^2}\,\frac{{\omega^2}/({k_z^2c_A^2})+3}{\left[1-{\omega^2}/({k_z^2c_A^2})\right]^2}\, .
\eeq

In dimensionless units normalized by $r_0$, \Eq{e:r0disp} becomes a quadratic equation for the dimensionless quantity ${\omega^2}/({k_z^2c_A^2})$:
\beq{e:biquadr}
\frac{3}{4}\left(\frac{c_A}{v_\phi(r_{0})}\right)^2\left(1-\frac{\omega^2}{k_z^2c_A^2}\right)^2-\frac{\omega^2}{k_z^2c_A^2}-3=0\,, \quad  v^2_\phi(r_0)\equiv\frac{GM}{r_0}=\frac{GMk_z}{\sqrt{-\tilde E_n}}\, .
\eeq
Its solution is as follows:
\beq{e:n2}
\omega^2=c_A^2k_z^2\,\left(1+ \frac{1\pm \sqrt{1+12\left[{c_A}/{v_\phi(r_{0})}\right]^2}}{({3}/{2})\,\left[{c_A}/{v_\phi(r_0)}\right]^2}\right)\,.
\eeq

Note that this dispersion equation involves the flow velocity at radius $r_0$ and does not explicitly depend on the boundaries of the flow where the boundary conditions are satisfied; the boundary conditions determine the set of eigenvalues $\tilde E_n$.

The critical magnetic field corresponding to the neutral mode $\omega^2=0$ is given by:
\beq{e:ncAcr}
\left(\frac{c_A}{v_\phi(r_0)}\right)_\mathrm{cr}^2=4\,,
\eeq
which can be rewritten as:
\beq{e:cacr}
(c_A)^2_\mathrm{cr}=\frac{4G\,M}{r_0}=
\frac{4G\,M\,k_z}{\sqrt{-\tilde E_n}}\,.
\eeq
Thus, in a sufficiently strong magnetic field, the shear flow is stabilized by the Lorentz force ($\omega^2=0$), as first noted by \citet{Velikhov59} for the case of flow in a narrow gap between two conducting cylinders.

Using \Eq{e:ncAcr}, we rewrite \Eq{e:n2} in the form:
\beq{e:n2cr}
\frac{\omega^2}{(c_A)^2_\mathrm{cr}k_z^2}=\myfrac{c_A}{(c_A)_\mathrm{cr}}^2\left(1+\frac{1\pm\sqrt{1+48\,({c_A}/{(c_A)_\mathrm{cr}})^2}}{6\,({c_A}/{(c_A)_\mathrm{cr}})^2}\right)\,.
\eeq
Below, only the unstable modes with $\omega^2<0$ are considered, corresponding to the minus sign in \Eq{e:n2cr}. For small $\left(\frac{c_A}{(c_A)_\mathrm{cr}}\right)^2\ll 1$,  \Eq{e:n2cr} can be approximated as
\beq{e:n2cr1}
\frac{\omega^2}{(c_A)^2_\mathrm{cr}k_z^2}\approx
    \left(\frac{c_A}{(c_A)_\mathrm{cr}}\right)^2\left(-3+48\,\left(\frac{c_A}{(c_A)_\mathrm{cr}}\right)^2\right)\,.
\eeq
In other words, $\omega^2\to 0$ as $c_A^2\to 0$. This behavior of $\omega^2$ differs from the result of local analysis, where the MRI occurs even for arbitrarily small (but nonzero!) background magnetic field (see, for example, \citealt{2015MNRAS.448.3697S}).

\begin{figure}
\begin{center}
        \includegraphics[width=0.8\textwidth]{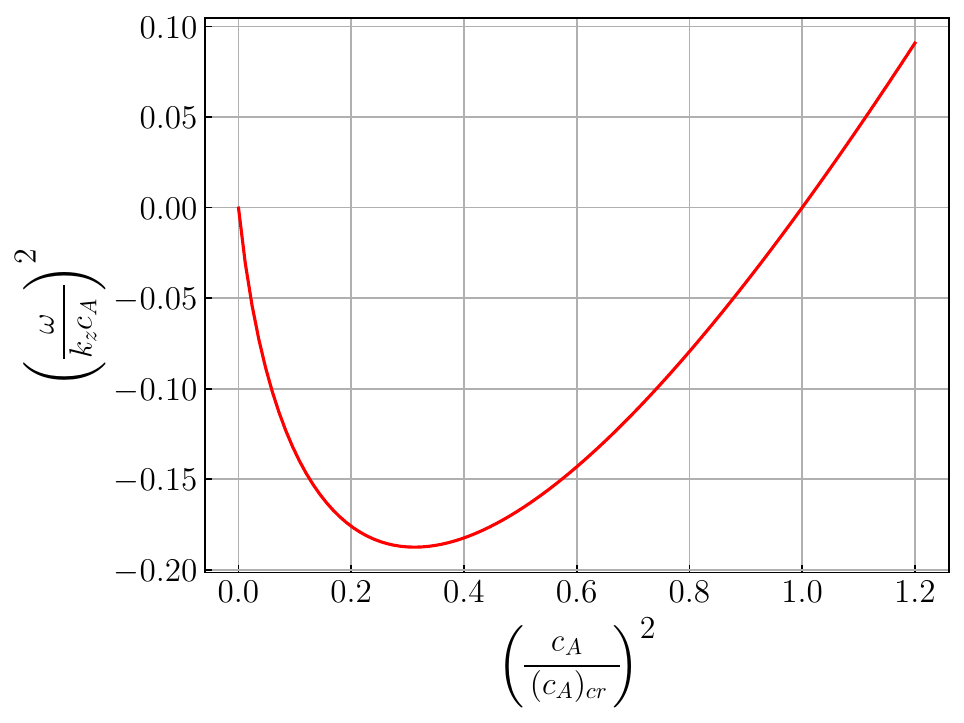}
        \caption{
        Solution of the dispersion equation (\ref{e:n2cr}) with a minus sign in front of the square root for unstable modes $\omega^2<0$.
        }
        \label{f:modes}
        \end{center}
    \end{figure}

It is evident that there exists a maximum growth rate of the MRI and its corresponding Alfv\'en speed (see Fig.~\ref{f:modes}):
\beq{e:max}
\myfrac{c_A}{(c_A)_\mathrm{cr}}^2_\mathrm{max}=\frac{5}{16},\qquad 
\frac{\omega^2}{(c_A)^2_\mathrm{cr}k_z^2}=-\frac{3}{16}\, .
\eeq

In real flows, there is always an inner radius $r_\mathrm{in}$. For sufficiently large outer radius $r_\mathrm{out}\gg r_\mathrm{in}$, it is possible to solve the problem exactly and find the eigenvalues $\tilde E_n=-k_z^2r_0^2$ (see Appendix A), from which the location of the zero point $r_0$ of the effective potential $U(r)$ can be calculated. In the quasi-classical approximation, the solution $\tilde E_n(x_\mathrm{in})$, where $x_\mathrm{in}=r_\mathrm{in}/r_0$, is expressed through the integral \eqref{e:BSx} (see Appendix A). Then, the position of $r_0$ in the flow can be calculated from the solution of the problem with a given $r_\mathrm{in}$ using the formula $r_0=r_\mathrm{in}/x_\mathrm{in}$. In the special case of small $k_z\approx 0$ (i.e., $\tilde E_n\approx 0$), there is an analytical solution (see \eqref{e:E0}, \eqref{e:disp0}):
\beq{}
r_0=r_\mathrm{in}\left\{\left[\frac{\pi}{\sqrt{3}}\left(n+\frac{3}{4}\right)+\frac{\pi}{2}\right]^2+1\right\}\,.
\eeq

In the case of $r_\mathrm{out}<r_0$, the potential between the boundaries does not change sign, and the problem of finding the eigenvalues of  \Eq{e:Sh1} is simplified (see Appendix B).

\subsection{Application to thin accretion disks}
\label{s:thin_ad}

In accretion disks, the wavelength of perturbations  $\lambda = 2\pi/k_z$ in the vertical coordinate must be smaller the disk semi-thickness $h$. This condition limits the allowed "energy levels." Indeed, at the turning point $x_1$, we have
\beq{}
\tilde E = -k_z^2r_0^2 = -4\pi^2\left(\frac{h}{\lambda}\right)^2\left(\frac{r_1}{h}\right)^2\frac{1}{x_1^2}\,.
\eeq
Therefore, the requirement
\beq{}
\left(\frac{\lambda}{h}\right)^2 = -\frac{4\pi^2}{x_1^2\tilde E}\left(\frac{r_1}{h}\right)^2 < 1
\eeq
gives the permissible energy levels that satisfy the following condition:
\beq{b:x12E}
x_1^2\,|\tilde E| > \frac{4\pi^2}{(h/r_1)^2}\,.
\eeq
For a typical value of the relative thickness of a geometrically thin disk, $h/r \lesssim 0.1$, in thin accretion disks, this yields $x_1^2|\tilde E| > 4\times 10^3$. Considering that for large $|\tilde E|\gg 1$, $x_1 \approx \left({3}/({4|\tilde E|})\right)^{1/3}$ (see \Eq{e:x1cardano}), the corresponding energy levels must have $|\tilde E| > 7\times 10^{10}$. Such levels are only possible for very small $x_\mathrm{in}\ll 1$ (see Appendix A, Fig.~\ref{f:psi_n}).

The critical magnetic field, above which the MRI is suppressed, can be expressed in terms of the Keplerian velocity at the inner boundary of the flow, $v_\phi(r_\mathrm{in})$:
\beq{e:cacrxin}
(c_A)_\mathrm{cr}^2=\frac{4GM}{r_0}=x_\mathrm{in}\frac{4GM}{r_\mathrm{in}}=4x_\mathrm{in}v_\phi^2(r_\mathrm{in})\,.
\eeq
It is easy to obtain an upper limit on the possible value of the dimensionless parameter $x_\mathrm{in}=r_\mathrm{in}/r_0$. Indeed, all possible energy levels can take values
\beq{}
-\tilde E_n=\frac{r_\mathrm{in}^2k_z^2}{x_\mathrm{in}^2}< -\tilde U(x_\mathrm{in})=\frac{3}{4} \frac{1}{x_\mathrm{in}^3}-\frac{3}{4} \frac{1}{x_\mathrm{in}^2}\,,
\eeq
from which
\beq{e:xinlim}
x_\mathrm{in}<\frac{1}{1+({4}/{3})\, r_\mathrm{in}^2\,k_z^2}\,.
\eeq
Note that this inequality gives an obvious upper limit $x_\mathrm{in}<1$ for $k_z\to 0$, while as shown in Appendix A, energy levels with $n=0$ exist only when $x_\mathrm{in}<0.8116$. 
Substituting inequality \eqref{e:xinlim} into \eqref{e:cacrxin}, we obtain an upper limit for the critical magnetic field:
\beq{e:cacrlim}
(c_A)_\mathrm{cr}^2<\frac{4v_\phi^2(r_\mathrm{in})}{1+({4}/{3})\, r_\mathrm{in}^2k_z^2}\,.
\eeq
Using the condition for thin disks with the wavelength of perturbations in the $z$-coordinate $\lambda/h=2\pi /(k_z h)<1$ and considering $h/r_\mathrm{in}<1$, we transform \eqref{e:cacrlim} into the form:
\beq{e:cacrd}
\frac{(c_A)_\mathrm{cr}}{v_\phi(r_\mathrm{in})} < 
\frac{\sqrt{3}}{2\pi}
\myfrac{h}{r_\mathrm{in}}\sqrt{\frac{\lambda}{h}}, \quad \frac{(c_A)_\mathrm{cr}}{c_s(r_\mathrm{in})} <
\frac{\sqrt{3}}{2\pi}\sqrt{\frac{\lambda}{h}}\,.
\eeq
(In the last inequality, we used the relation between the Keplerian velocity and the sound speed at the inner boundary of the thin accretion disk $c_s(r_\mathrm{in})=v_\phi(r_\mathrm{in})(h/r_\mathrm{in})$.)

Thus, in thin accretion disks, MRI perturbations with a wavelength $\lambda<h$ in the $z$-coordinate  can be significantly suppressed.

Depending on the type of accretor (regular stars or relativistic compact objects), the inner radius of the flow can vary greatly from a few tens to millions of kilometers, while the outer radius is determined by the physical situation, such as being limited by the size of the Roche lobe in the case of binary systems. Since in thin disks $h/r\sim 0.05$, the upper limit of the critical magnetic field that suppresses MRI in thin disks \eqref{e:cacrd} is $\sim 1/\sqrt{r_\mathrm{in}}$ and significantly differs for different types of accretors.

Previous studies on the critical magnetic field for suppressing MRI in the context of global analysis with different boundary conditions can be found, for example, in \cite{1992GApFD..66..223P,2015MNRAS.453.3257L}. {\cite{1992GApFD..66..223P} derived the critical magnetic field using the energy method. Only an upper estimate was obtained, according to which (equation (41) in the cited work) the critical magnetic field in thin disks is determined by the Alfv\'en velocity, which is equal to the Keplerian velocity of the flow multiplied by the square root of the ratio of the disk's semi-thickness to its radius. 
It should also be noted that the critical Alfv\'en velocity in the local analysis by \cite{1991ApJ...376..214B} is, to within a multiplicative factor, equal to the Keplerian velocity multiplied by the ratio of the wavelength of a perturbation in the $z$-coordinate to the radius. 
As can be seen from the comparison of these statements with \eqref{e:cacrd}, our result provides a much more definite value for the critical field.}

\section{MRI with radially dependent Alfv\'en velocity}

Until now, our consideration of MRI has been limited to the case of a constant background Alfv\'en velocity. However, in real astrophysical situations, the Alfv\'en velocity should decrease with radius at least as fast as the angular velocity of the flow or even faster. Below, we will consider two cases: (1)  $c_A^2$ depends on radius as $v_\phi^2\propto 1/r$ due to the variable density with a constant background magnetic field, and (2)  $c_A^2\propto 1/r^q$ for a variable background magnetic field with the density constant over radius.

\subsection{Case of constant background magnetic field} \label{s:econst}
The square of the Alfvén velocity can decrease linearly with radius when the background magnetic field is constant, but the density of the flow increases \textit{linearly with radius}. It should be noted that the density varying with radius still implies the use of the continuity equation in the form of \Eq{iurz}. Let us rewrite \Eq{e:cont} as follows: 
\[
\frac{1}{\rho} \left(\frac{\partial \rho}{\partial t}+\bm{u}\,\bm{\nabla}\rho\right) + \bm{\nabla}\,\bm{u} = \frac{{\rm D}\ln\rho}{{\rm D}t}+\bm{\nabla}\,\bm{u} = 0.
\]
Now, locally, the density $\rho$ varies (Eulerian derivative $\partial \rho/\partial t\neq 0$), but the substantial (Lagrangian) derivative of density with time ${\rm D}\rho/{\rm D}t=\partial \rho/\partial t+\bm{u}\,\bm{\nabla} \rho=0$, ${\rm D}\ln\rho/{\rm D}t=0$. Thus, for an incompressible fluid in the case of density varying with coordinates, the continuity equation becomes $\bm{\nabla}\,\bm{u}=0$ \citep{landau-lifshitz-hydro,1990Acheson}. 
Therefore, for a variable background density $\rho_0$, equations (\ref{iurz}--\ref{ibz}) for linear perturbations remain unchanged. However, instead of the potential \Eq{e:UK} with $c_A^2=\mathrm{const}$, the effective potential takes the form
\begin{eqnarray}
\label{e:UK1}
&U_e=\frac{3}{4}\frac{1}{r^2}-\frac{GM}{r^3c_A^2}
\frac{{\omega^2}/({k_z^2c_A^2})+3}{\left[1-{\omega^2}/({k_z^2c_A^2})\right]^2}=\nonumber\\
& =\frac{3}{4}\frac{1}{r^2}-\frac{GM}{r^3}
\frac{{\omega^2}/{k_z^2}+3\epsilon v_\phi^2(r)}{\left[\epsilon v_\phi^2(r)-{\omega^2}/{k_z^2}\right]^2}\,.
\end{eqnarray}
Here $\epsilon\equiv c_A^2/v_\phi^2=\mathrm{const}$ is a parameter. The potential (\ref{e:UK1}) has the same characteristic points as in the previous case (see Section \ref{s:constantCA}): $r_0$, where $U_e(r_0)=0$, the turning point $r_1$, where $E-U_e(r_1)=0$, and the inner radius of the flow $r_\mathrm{in}$.

The potential $U_e$ becomes zero at the point 
\beq{e:r0e}
r_0=\frac{4}{3}\,G\,M\,
\frac{{\omega^2}/{k_z^2}+3\epsilon v_\phi^2(r_0)}{\left[\epsilon v_\phi^2(r_0)-{\omega^2}/{k_z^2}\right]^2}
\,.
\eeq
From the last equation, we obtain the dispersion equation $\omega(k_z)$:
\beq{e:dispe}
\frac{3}{4}\,\epsilon\,\left(1-\frac{\omega^2}{k_z^2\epsilon\, v_\phi^2(r_0)}\right)^2-\frac{\omega^2}{k_z^2\,\epsilon\, v_\phi^2(r_0)}-3=0
\eeq
with the solution
\beq{e:n2e}
\omega^2=\epsilon\, v_\phi^2(r_0) \,k_z^2\,\left(1+ \frac{1\pm \sqrt{1+12\epsilon}}{{3\epsilon}/{2}}\right)\,.
\eeq
The unstable MRI mode corresponds to the minus sign before the square root. As in Section \ref{s:de}, the critical magnetic field corresponds to the neutral mode $\omega^2=0$:
\beq{}
\epsilon_\mathrm{cr}=\myfrac{c_A(r_0)}{v_\phi(r_0)}^2=4\,.
\eeq
For small $\epsilon\ll 1$, 
\beq{e:n2e1}
\omega^2\approx \epsilon \,v^2_\phi(r_0)\, k_z^2\,\left(-3+\epsilon\right)\,.
\eeq
It can be seen that \Eq{e:n2e} yields the same maximum growth rate of MRI and the corresponding value of the Alfv\'en velocity as before (cf.~\Eq{e:max}):
\beq{e:maxe}
\epsilon=\myfrac{c_A(r_0)}{v_\phi(r_0)}_\mathrm{max}^2=\frac{5}{4}\,,\qquad \omega^2_\mathrm{max}=-\frac{3}{4}\, k_z^2\,v_\phi^2(r_0)\,.
\eeq
Clearly,  the critical Alfv\'en velocity is the same as derived earlier (see \Eq{e:ncAcr}).

Now let us consider the turning point $r_1$ in the potential $U_e$, determined by the equation $-k_z^2-U_e(r_1)=0$. In dimensionless variables $\tilde E=-k_z^2\,r_0^2$ and $x=r/r_0$, the equation for $x_1$ reads:
\beq{e:x1}
\tilde E-\tilde U_e =\tilde E-\frac{3}{4}\frac{1}{x_1^2}+\frac{1}{\epsilon x_1^2}\frac{3+K(\epsilon)}{[1-x_1\,K(\epsilon)]^2}=0\,,
\eeq
where
\beq{e:Ke}
K(\epsilon)=\frac{\omega^2}{k_z^2\,c_A^2(r_0)}=1+ \frac{1- \sqrt{1+12\epsilon}}{{3\epsilon}/{2}}\,.
\eeq
It should be noted that the "attractive" part of the potential in \Eq{e:x1} depends on the parameter $\epsilon$. Also, in this case, for small $x\ll 1$, the "repulsive" potential behaves as $1/x^2$ compared to $1/x^3$ in the case of \Eq{e:Ux}. \Eq{e:x1} is a fourth-degree equation for $x_1$ (in contrast to the cubic equation \Eq{e:ntp} for $x_1$ in the case of constant $c_A^2$). 
For small $x_1\ll 1$, the solution takes the form:
\beq{e:x10}
x_1^2\approx-\frac{3}{\tilde E}\left(
\frac{1}{4}+\frac{1}{\epsilon}\right)
\eeq
(cf. \Eq{e:x1cardano} for constant Alfv\'en velocity).

\subsection{Case of radially dependent background magnetic field}\label{s:CAq}

If the background magnetic field depends on the radius, \hbox{$\bm{B}=(0,0,B_z(r))$}, the linearized MHD equations become more complex. The radial and vertical components of the Euler equations \ref{iur} and \ref{iuz} are now expressed as follows, respectively:
\beq{iur1}
\ii \omega u_r-2\Omega 
u_\phi=-{\frac{1}{\rho_0}\frac{\partial p_1}{\partial r}}
-\frac{1}{4\pi \rho_0}\left[\dfrac{(b_zB_z)}{r}+\ii k_zb_rB_z\right]\,,
\eeq
\beq{iuz1}
\ii \omega u_z=\ii k_z\frac{p_1}{\rho_0} + \frac{1}{4\pi \rho_0}b_r\dfrac{B_z}{r}\,.
\eeq
The vertical component of the induction equation \eqn{ibz} turns into
\beq{ibz1}
\mathrm{i}\omega b_z=-\frac{1}{r}\dfrac{(ru_rB_z)}{r}\,.
\eeq
Following a similar procedure as in Section \ref{s:constantCA}, after some algebraic manipulations, we obtain the equation for $b_r$ in the form:
\begin{eqnarray}
\label{e:br1}
\displaystyle
&\left(c_A^2-\frac{\omega^2}{k_z^2}\right)\,\left[
\frac{\partial^2b_r}{\partial r^2}+\frac{\partial }{\partial r}\myfrac{b_r}{r}-b_rk_z^2\right]+\nonumber\\
&\frac{
2\,\Omega\,\left[{\omega^2\,\varkappa^2}/({2\,\Omega\,k_z^2})-c_A^2\,r\,({{\rm d}\Omega}/{{\rm d}r})\right]}{c_A^2-{\omega^2}/{k_z^2}}\, b_r=\nonumber\\
&\frac{\partial c_A^2}{\partial r}\,\Bigg\{
-\frac{b_r}{2r}-\frac{1}{2}\frac{\partial b_r}{\partial r}+\frac{\omega^2}{k_z^2c_A^2} \times\nonumber\\
&\times\left[
-\frac{b_r}{2r}-\frac{\partial b_r}{\partial r}-
\frac{1}{2}\frac{\partial^2c_A^2/\partial r^2}{\partial c_A^2/\partial r}\,b_r+\frac{3}{4c_A^2}\,\frac{\partial c_A^2}{\partial r}\,b_r
\right]
\Bigg\}
\,.
\end{eqnarray}
The r.h.s. of the above equation vanishes for $c_A^2=$const. 
Here, we will consider the case of a power-law dependence of the Alfv\'en velocity on radius. Substituting $c_A^2\propto r^{-q}$ into \Eq{e:br1}, we obtain, for Keplerian flow ($\varkappa^2=\Omega^2 = GM/r^3$), the following expression:
\begin{eqnarray}
\label{e:brq}
&\frac{\partial^2b_r}{\partial r^2}+\left[1-q\,\frac{{c_A^2}/{2}+{\omega^2}/{k_z^2}}{c_A^2-\omega^2/k_z^2}
\right]\,
\frac{\partial }{\partial r}\myfrac{b_r}{r}+\nonumber\\
&+b_r\Bigg\{-k_z^2+
\frac{G\,M}{r^3}\,
\frac{{\omega^2}/{k_z^2}+3c_A^2}{\left(c_A^2-{\omega^2}/{k_z^2}\right)^2}+\nonumber\\
&+\frac{q}{r^2}
\frac{\left(-{q}/{4}-1\right)\,{\omega^2}/{k_z^2}-c_A^2}{c_A^2-{\omega^2}/{k_z^2}}
\Bigg\}=0\, .
\end{eqnarray}

\Eq{e:brq} can be rewritten as
\begin{eqnarray}
&b_r''+g(r)\, b_r'+f(r)=0\,,\\
&g(r)=\frac{1}{r}\frac{1-{q}/{2}-(q+1)\,{\omega^2}/\left({k_z^2c_A^2}\right)}{1-{\omega^2}/\left({k_z^2c_A^2}\right)}\, ,\\
&f(r)=-k_z^2+\frac{G\,M}{c_A^2r^3}\,
\frac{{\omega^2}/\left({k_z^2c_A^2}\right)+3}{\left[1-{\omega^2}/\left({k_z^2c_A^2}\right)\right]^2}-\nonumber\\
&-\frac{g(r)}{r}+\frac{q}{r^2}
\frac{\left(-{q}/{4}-1\right)\,{\omega^2}/\left({k_z^2c_A^2}\right)-1}{1-{\omega^2}/\left({k_z^2c_A^2}\right)} \, .
\end{eqnarray}
The first derivative can be excluded in a standard way by the substitution $z=b_r\exp\left[-{1}/{2}\int^rg(s)\,{\rm d}s\right]$, yielding the equation 
\beq{e:z}
z''+\left(f-\frac{g'}{2}-\frac{g^2}{4}\right)z=0\,.
\eeq
Substituting the functions $f$ and $g$ into \Eq{e:z}, we obtain the equation:
\begin{eqnarray}
&
z''+\Bigg\{
-k_z^2-\frac{3}{4}\frac{1}{r^2}
-\frac{q}{r^2}\,\frac{{q}/{16}+{1}/{2}-\left(q/4+1/2\right)\,{\omega^2}/\left(k_z^2c_A^2\right)}{\left[1-{\omega^2}/\left({k_z^2c_A^2}\right)\right]^2} +\nonumber\\
&+\,\frac{G\,M}{c_A^2r^3}\,
\frac{{\omega^2}/\left({k_z^2c_A^2}\right)+3}{\left[1-{\omega^2}/\left({k_z^2c_A^2}\right)\right]^2}
\Bigg\}\,z=0\,.
\label{e:zq}
\end{eqnarray}
Clearly, it reduces to \Eq{e:shred} for $q=0$. 
for the case of a constant magnetic field. For the case when $q\ne 0$, the effective potential in the Schr\"odinger-like \Eq{e:zq} reads:
\begin{eqnarray}
\label{e:Uz}
&U_q=\frac{3}{4}\frac{1}{r^2}
+\frac{q}{r^2}\,\frac{{q}/{16}+{1}/{2}-\left({q}/{4}+{1}/{2}\right)\,{\omega^2}/\left({k_z^2c_A^2}\right)}{\left[1-{\omega^2}/\left({k_z^2c_A^2}\right)\right]^2} - \nonumber \\
&-\,\frac{GM}{c_A^2r^3}\,
\frac{{\omega^2}/\left({k_z^2c_A^2}\right)+3}{\left[1-{\omega^2}/\left({k_z^2c_A^2}\right)\right]^2}\, .
\end{eqnarray}
Foe each value of $q$, the potential $U_q$ becomes zero at:
\begin{eqnarray}
\label{e:Ur0}
&r_0=\frac{GM}{c_A^2} \times \nonumber \\
&\times\frac{{\omega^2}/\left({k_z^2c_A^2}\right)+3}{({3}/{4})\left[1-{\omega^2}/\left({k_z^2c_A^2}\right)\right]^2-\left({q^2}/{4}+{q}/{2}\right) {\omega^2}/\left({k_z^2c_A^2}\right)+\left({q^2}/{16}+{q}/{2}\right)}\, .\nonumber \\
&
\end{eqnarray}
The dispersion equation at the point $r_0$ reads:
\begin{eqnarray}
\label{e:deq}
&\frac{\omega^2}{k_z^2c_A^2(r_0)}=1+\frac{q}{3} +
\frac{q^2}{6}
+\frac 23\,\myfrac{v_\phi(r_0)}{c_A(r_0)}^2\pm\nonumber\\
&\pm  \sqrt{\left(\frac{q^2}{6}+\frac{q}{3}+\frac 23\myfrac{v_\phi(r_0)}{c_A(r_0)}^2\right)^2+\frac{16}{3}\myfrac{v_\phi(r_0)}{c_A(r_0)}^2+ \frac{q^2}{4}}.
\end{eqnarray}
The unstable MRI mode corresponds to the minus sign before the square root.
The zero mode $\omega^2=0$ occurs at the critical magnetic field
\beq{e:ncAcrq}
\epsilon_{cr}\equiv\myfrac{c_A(r_0)}{v_\phi(r_0)}_\mathrm{cr}^2=\frac{48}{12+8q+q^2}\,.
\eeq

\begin{figure}
 \begin{center}
        \includegraphics[width=0.8\textwidth]{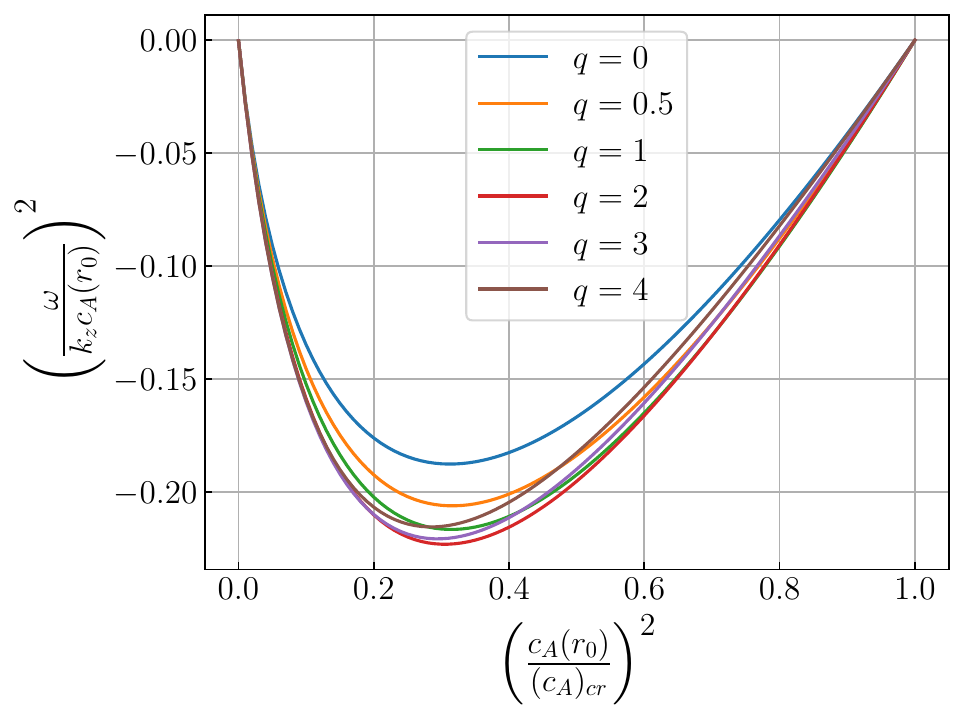}
        \caption{
        MRI dispersion curves described by  \Eq{e:deq1} for various values of $q$. In the limit of $c_A/(c_A)_\mathrm{cr}\ll 1$, the dispersion equation takes the form of \Eq{e:n2cr1} for all values of $q$.
        }
        \label{f:modesq}
        \end{center}
    \end{figure}

 Using the identity
 \beq{e:ncAcrq1}
 \myfrac{v_\phi}{c_A}^2=\myfrac{v_\phi}{(c_A)_\mathrm{cr}}^2\myfrac{(c_A)_\mathrm{cr}}{c_A}^2\, ,
 \eeq
 we rewrite the dispersion equation \ref{e:deq} in the form similar to \Eq{e:n2cr}, as a function of a dimensionless value $\left({(c_A)_\mathrm{cr}}/{c_A}\right)^2$:
\begin{eqnarray}
\label{e:deq1}
&\frac{\omega^2}{(c_A)^2_\mathrm{cr}k_z^2}=\myfrac{c_A}{(c_A)_\mathrm{cr}}^2\left[1+\frac{q}{3} +
\frac{q^2}{6}
+\frac 23\,\myfrac{v_\phi}{c_A}^2\right. \pm\nonumber\\
&\left.\pm  \sqrt{\left(\frac{q^2}{6}+\frac{q}{3}+\frac 23\myfrac{v_\phi}{c_A}^2\right)^2+\frac{16}{3}\myfrac{v_\phi}{c_A}^2+ \frac{q^2}{4}}\,\right].
\end{eqnarray}
(see Fig. \ref{f:modesq}). In \Eq{e:deq1},  value $\left({v_\phi}/{c_A}\right)^2$ can be expressed in terms of Eqs.~\eqref{e:ncAcrq} and \eqref{e:ncAcrq1}.
  
 \begin{figure*}
        \includegraphics[width=0.49\textwidth]{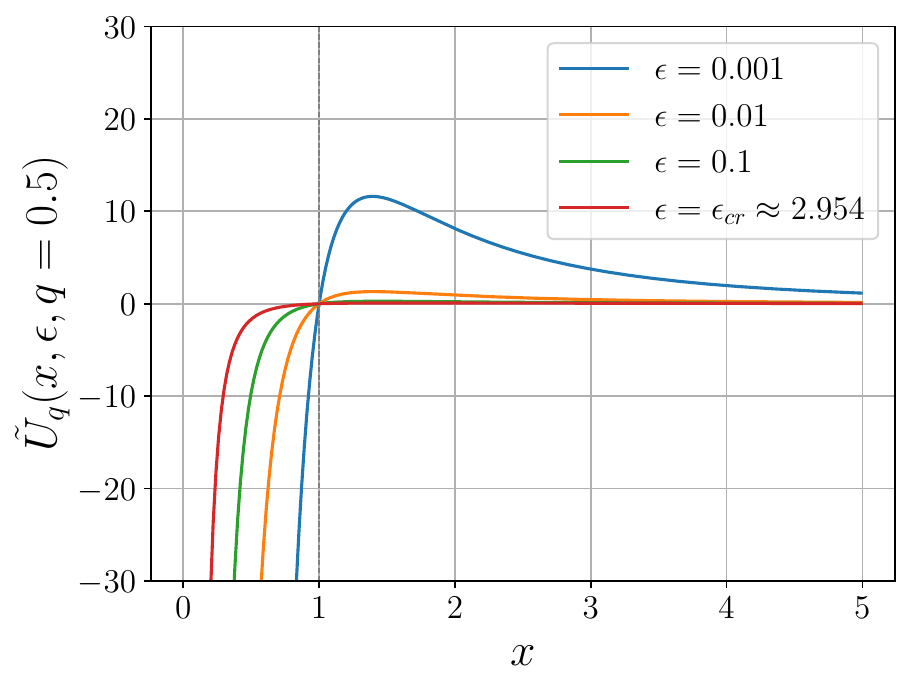}
        \includegraphics[width=0.49\textwidth]{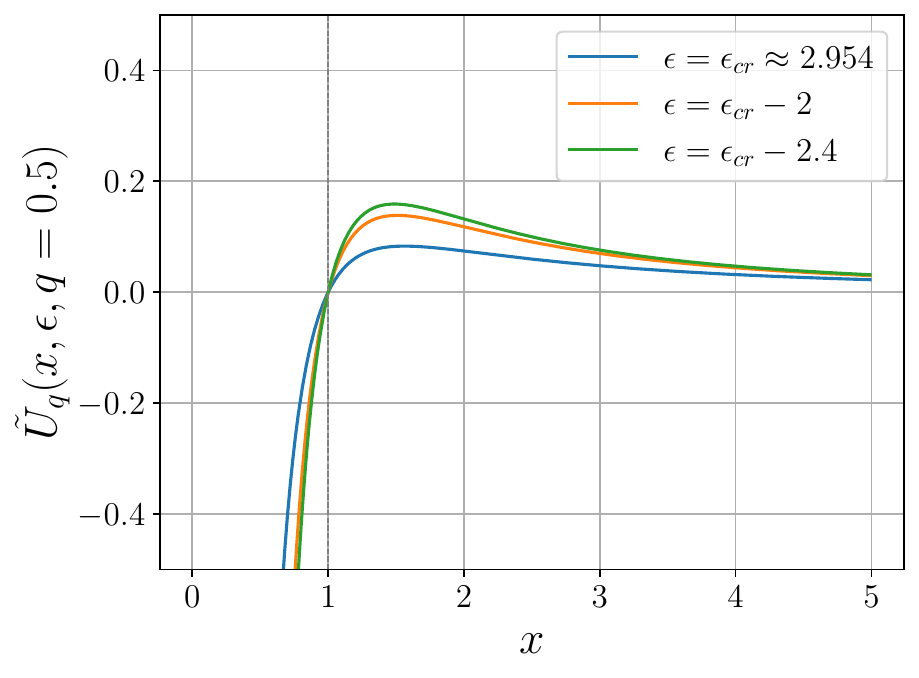}
        \caption{The effective potential $\tilde U_q(x,\epsilon,q)$ described by \Eq{e:uq} for $q=0.5$ and various values of the parameter $\epsilon=c_A^2(r_0)/v_\phi^2(r_0)$. The critical value $\epsilon_\mathrm{cr}$ is determined from  \Eq{e:ncAcrq}.
        }
        \label{f:Uq05}
    \end{figure*}

\begin{figure*}
        \includegraphics[width=0.49\textwidth]{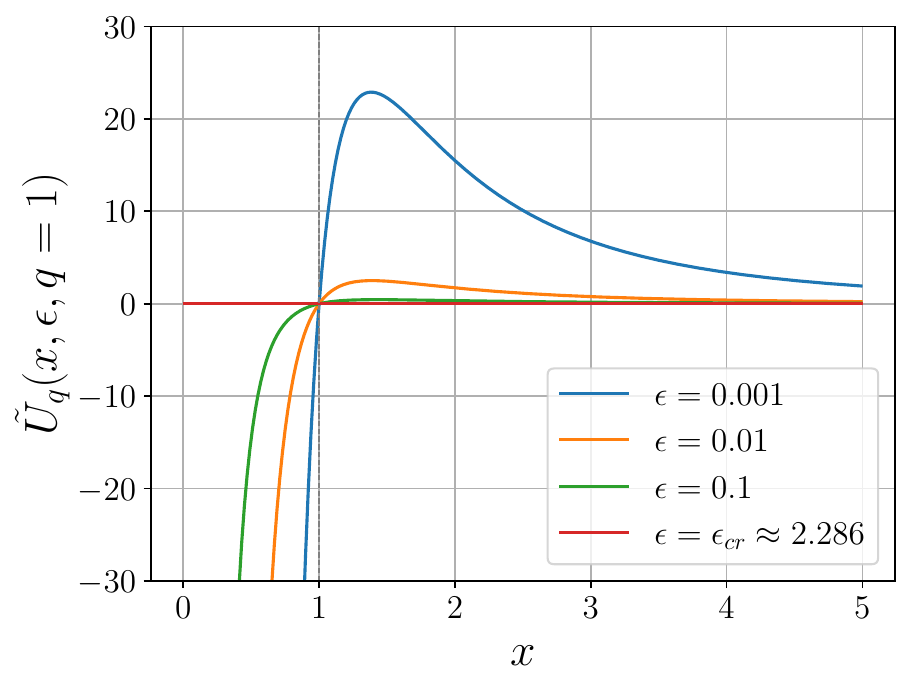}
        \includegraphics[width=0.49\textwidth]{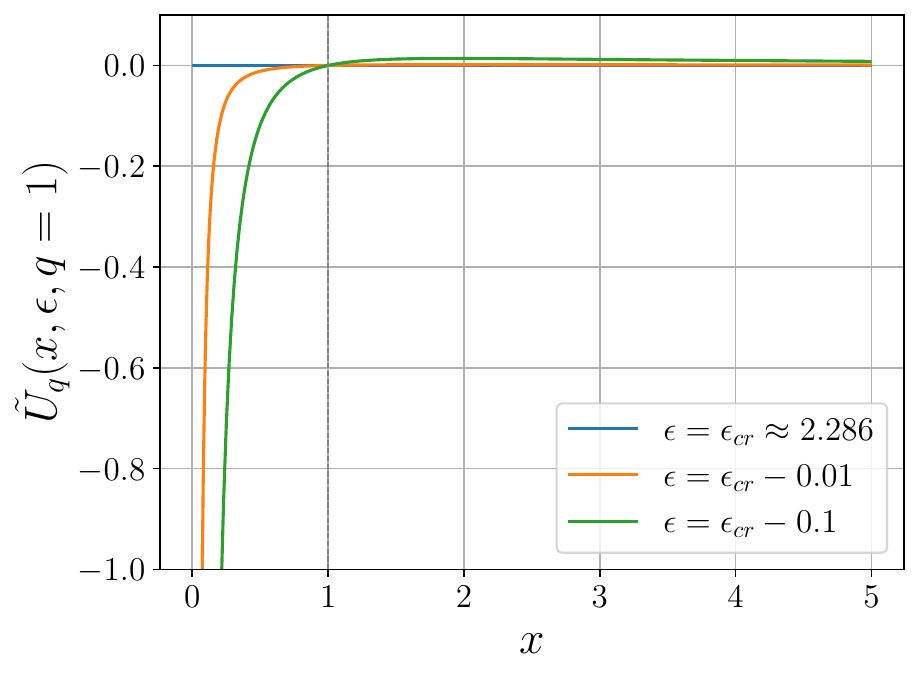}
        \caption{The same as in Fig. \ref{f:Uq05}, for $q=1$.}
        \label{f:Uq1}
    \end{figure*}
    
\begin{figure*}
        \includegraphics[width=0.49\textwidth]{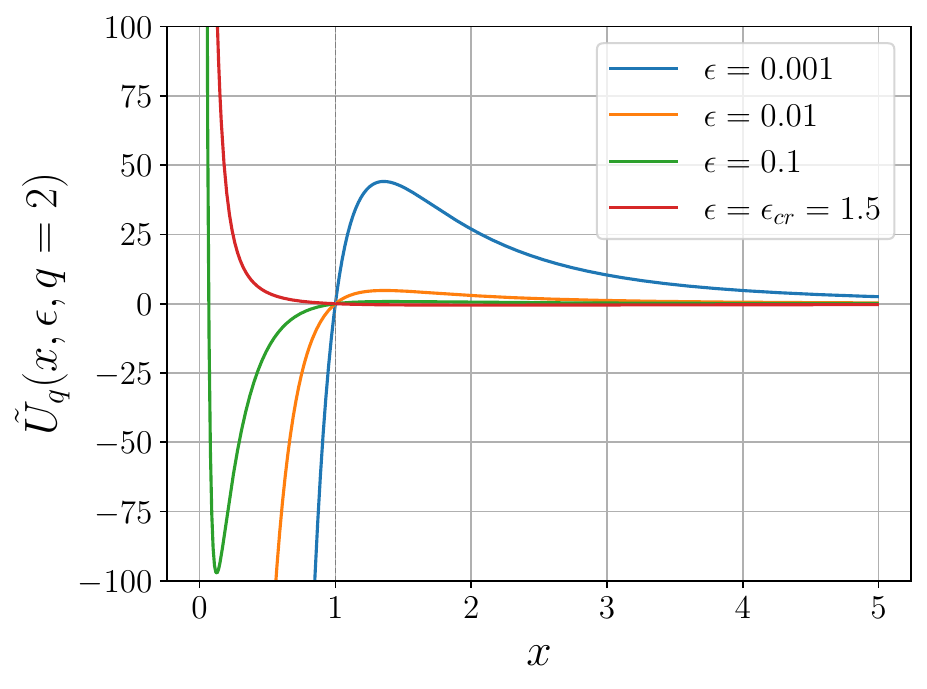}
        \includegraphics[width=0.49\textwidth]{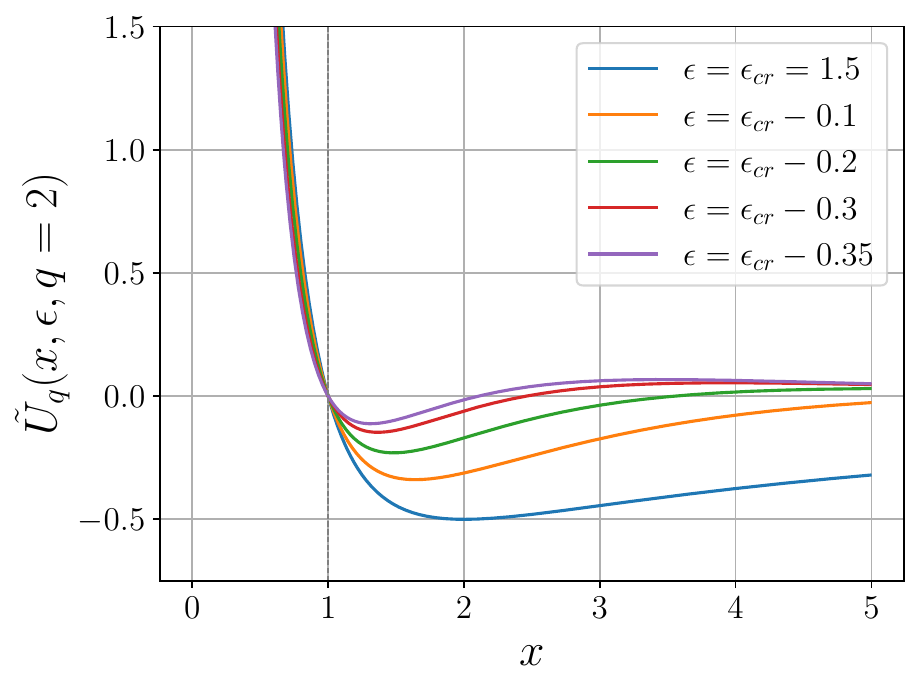}
        \caption{The same as in Fig. \ref{f:Uq05}, for $q=2$.}
        \label{f:Uq2}
    \end{figure*}
    
\begin{figure}
        \includegraphics[width=0.49\textwidth]{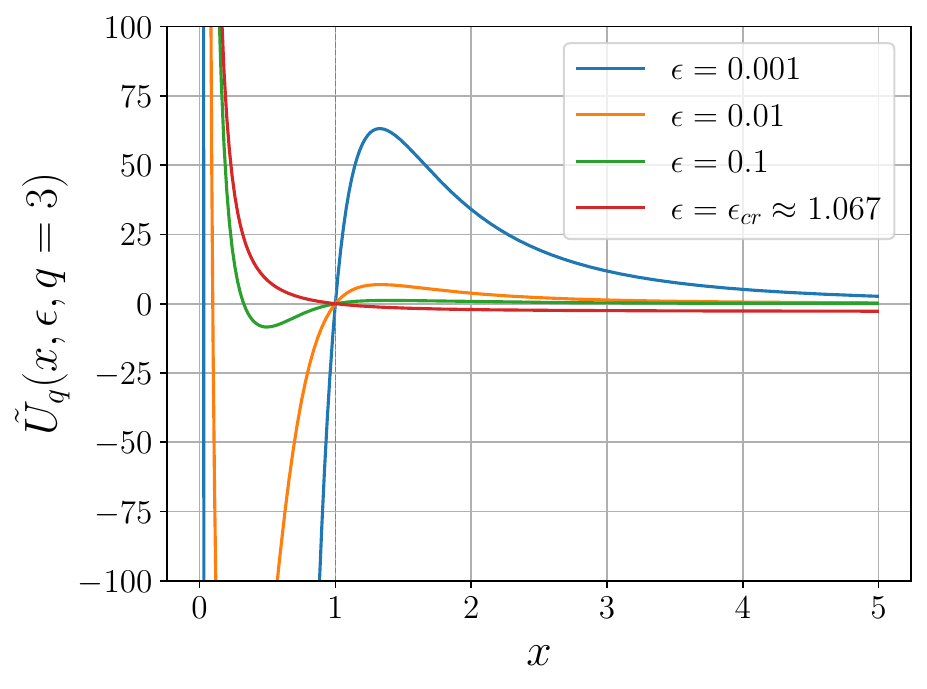}
        \includegraphics[width=0.49\textwidth]{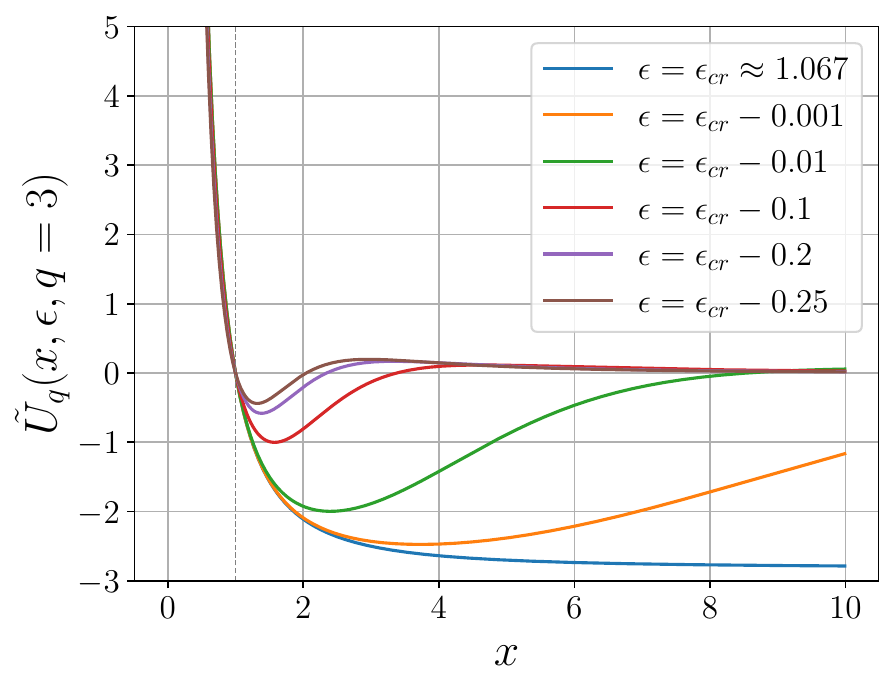}
        \caption{
        The same as in Fig. \ref{f:Uq05}, for $q=3$.
        }
        \label{f:Uq3}
    \end{figure}
    
\begin{figure}
        \includegraphics[width=0.49\textwidth]{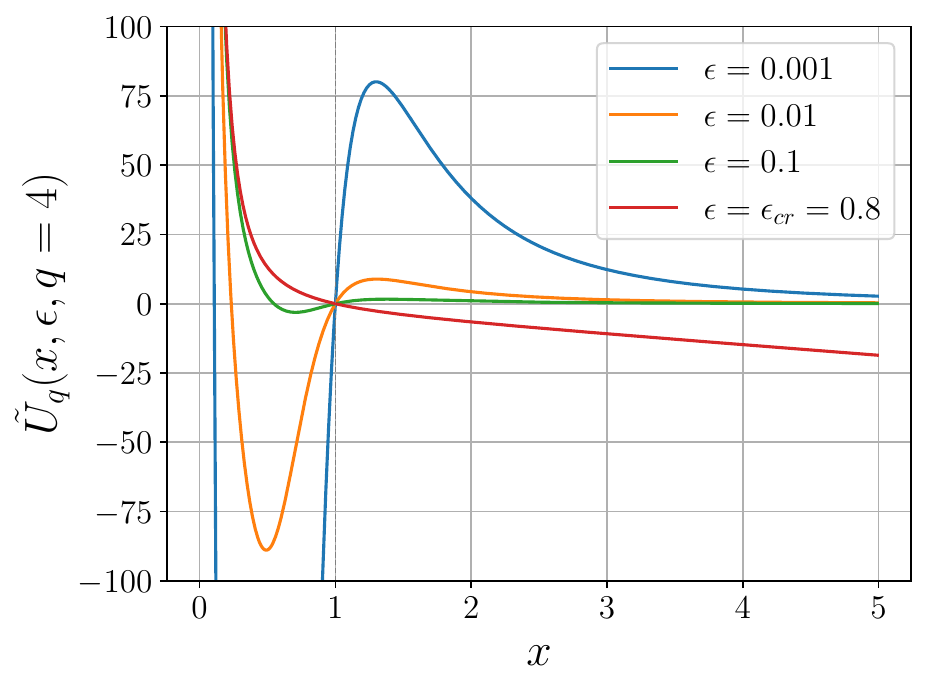}
        \includegraphics[width=0.49\textwidth]{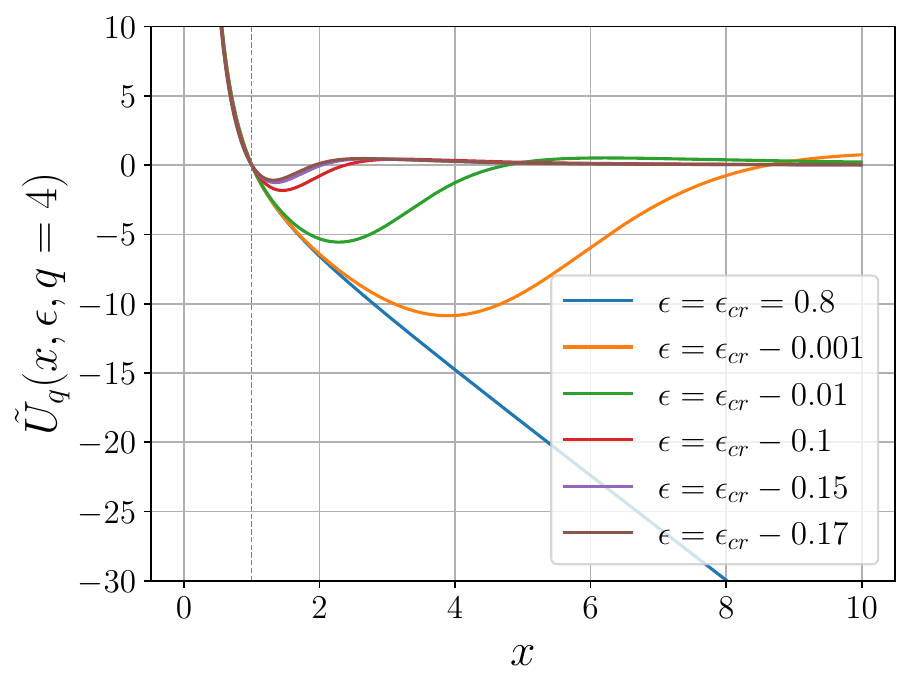}
        \caption{
        The same as in Fig. \ref{f:Uq05}, for $q=4$.
        }
        \label{f:Uq4}
    \end{figure}

Let us use, as above, the dimensionless variables $x=r/r_0$, $c_A^2=c_A^2(r_0)x^{-q}$, parameter $\epsilon=c_A^2(r_0)/v_\phi^2(r_0)$ and denote the r.h.s. of \Eq{e:deq} ${\omega^2}/\left({k_z^2c_A^2(r_0)}\right)=K(\epsilon)$. Then the dimensionless potential $\tilde U_q=U_qr_0^2$ is as follows
\begin{eqnarray}
\label{e:uq}
&\tilde U_q(x,\epsilon,q)=\frac{1}{x^2}\, \left(
\frac{3}{4}+\frac{q^2/16+{q}/{2}-K(\epsilon)\,x^q\,({q^2}/{4}+{q}/{2})}{(1-K(\epsilon)x^q)^2}\right)-\nonumber\\
&\frac{1}{x^{3-q}}\, \frac{3+K(\epsilon)\,x^q}{\epsilon\,(1-K(\epsilon)\,x^q)^2}\,.
\end{eqnarray}
It should be noted that in the case of $q\neq 0$, both the "repulsive" and "attractive" parts of the effective potential depend on $\epsilon$.

Let us discuss some features of the potential $\tilde U_q$.
\begin{enumerate}
    \item 
    By construction, $\tilde U_q(1,\epsilon,q)=0$ for any $\epsilon, q$. \item
    For $\epsilon\to \epsilon_\mathrm{cr}$,  see \Eq{e:ncAcrq},  $K(\epsilon)\to 0$:
    \begin{equation}
    \label{e:uq1}
    \tilde U_q(x,\epsilon\to\epsilon_\mathrm{cr},q)=\left(
    \frac{3}{4}+\frac{q^2}{16}+\frac{q}{2}\right)
    \left(\frac{1}{x^2}-\frac{1}{x^{3-q}}\right)\,.
    \end{equation}    
The potential $\tilde U_q(x,\epsilon\to\epsilon_\mathrm{cr},q=1)=0$ for any $x$. For $q<1$, $\tilde U_q$ changes sign from negative to positive at $x=1$, and vice versa for $q>1$. 
    \item For $q<1$, there exists only one zero point of the potential: $\tilde U_q(x=1, \epsilon,q<1)=0$. 
\item For $q>1$, the second zero point $x_2$ appears at $x<1$ with $\lim_{\epsilon\to 0}x_2=0$ and $\lim_{x\to 0}\tilde U_q(x,\epsilon,q)\to +\infty$. That is, the potential reaches a minimum at a certain point $x_{\min}$: $x_2<x_{\min}<1$. 
\end{enumerate}

These features of the effective potential $\tilde U_q$ for different values of $q$ and $\epsilon$ are shown in Fig.~\ref{f:Uq05}-\ref{f:Uq4}. It can be seen that for $q\leq 1$, the effective potential $\tilde U_q$ has the same shape as for $q=0$ (see Fig.~\ref{f:1}), and the analysis of the MRN modes remains unchanged. 
For $q>1$, a minimum of the effective potential appears with $x_2<x_{\min}<1$, and the unstable energy levels should be sought between the two turning points for the corresponding effective energies $\tilde E=-k_zr_0^2$. 
In contrast to the $q=0$ case, there exists a minimum energy $\tilde E_{\min}=\tilde U_q(x_{\min},\epsilon,q)$.

It is important to note that for cases with $q>1$ and $x_\mathrm{in}\ll 1$, there can be two turning points in the effective potential $\tilde U_q$ for the energy levels $\tilde E=-k_z^2r_0^2$ ($[x_1]_{\min}$ and $[x_1]_{\max}$). It is also possible that $\tilde E<[\tilde{U}_q]_{\min}$, and stable negative energy levels (corresponding to MRN modes) do not exist. If the dimensionless radius of the inner disk $x_\mathrm{in}$ falls between the roots of the effective potential $\tilde U_q$, then a free boundary of the flow occurs at the point $x_\mathrm{in}$, as in the previously considered $q=0$ case.

Thus, the analysis of the shape of the effective potential $\tilde U_q$ indicates the need to take it into account when considering global MRN in specific physical cases.

\section{Summary and discussion}
\label{s:discussion}

\subsection{Non-local modal analysis with constant background magnetic field.}
We have revisited the development of MRI in Keplerian flows of ideal fluid. We have shown that taking into account radial nonlocality in the analysis of perturbations in the form of $f(r)\,e^{\ii (\omega t-k_zz)}$ leads to the appearance of the term $-({3}/{4})\,({1}/{r^2})$ in \Eq{e:shred} for small perturbations. 
The equation for small perturbations takes the form of a stationary Schrödinger equation with an effective potential that determines (in the form of a dispersion equation) the region of negative $\omega^2$ values--the region of MRI growth. The neutral mode $\omega^2=0$ corresponds to the critical magnetic field (\ref{e:cacr}). 
The critical field in terms of the Alfv\'en velocity $c_A^2$ is given by $(c_A)_\mathrm{cr}^2=4GM/r_0$, where $r_0$ is the zero point of the potential $U$ (\ref{e:UK}) in \Eq{e:shred}.

The second important result is the significant reduction (compared to the local analysis) of the MRI instability increment for the case of $c_A^2\ll (c_A)_\mathrm{cr}^2$ (see \Eq{e:n2cr1}). Indeed, neglecting the term $-\frac{3}{4}\frac{1}{r^2}$ in the equation \ref{e:shred} and replacing the derivative $\frac{\partial}{\partial r}\to -i k_r$, the equation \ref{e:shred} transforms into a fourth-degree algebraic equation with the solution
\beq{e:algebra}
\omega^2=\myfrac{k_z}{k}^2
\left[c_A^2k_z^2+\frac{\Omega^2}{2}-\sqrt{\frac{\Omega^4}{4}+4\Omega^2c_A^2k^2}\,\right]
\eeq
(here $k^2=k_r^2+k_z^2$). The maximal instability increment in this case is independent of the magnetic field strength: 
\beq{e:omegamax}
\omega^2_\mathrm{max}=-\frac{9}{16}\myfrac{k_z}{k}^2\Omega^2\,.
\eeq

We stress the difference between the local result (\ref{e:omegamax}) and the nonlocal result (\ref{e:max}). 
In the nonlocal approach, a critical magnetic field that suppresses MRI appears, which was absent in the local modal analysis (see, for example, \cite{1991ApJ...376..214B,2018ASSL..454..393S}). 
In the global analysis, the maximum increment of MRI is achieved at a specific value of the Alfv\'en velocity, $(c_A/(c_A)_{\rm cr})^2=5/16$ (see Fig.~\ref{f:modes} and \Eq{e:max}). The fact that MRI arises only in sufficiently weak magnetic fields was already noted in the seminal work by \citet{1991ApJ...376..214B} (see also, for example, subsequent studies \cite{2007MNRAS.375..177S,2013ApJ...769...76B}, etc.). For small magnetic fields, $c_A\ll (c_A)_\mathrm{cr}$, the instability increment is suppressed by a factor of $3\,(c_A/(c_A)_\mathrm{cr})$, as shown by \Eq{e:n2cr1}.

We emphasize that, as shown by numerical analysis (see Appendix A), in "shallow" potential wells with the dimensionless inner radius of the flow $x_\mathrm{in}=r_\mathrm{in}/r_0>0.8$, there are no stationary eigenmodes \ref{e:shred}, i.e., MRI is absent there. 
For $x_\mathrm{in}=r_\mathrm{in}/r_0<0.8$, stationary levels arise, corresponding to very small $k_z$ and large perturbation wavelengths. 
The number of stationary levels increases as $x_\mathrm{in}$ decreases. 
In thin Keplerian accretion disks, perturbations with wave lengths $\lambda=2\pi/k_z$ smaller than the disk's semi-thickness $h$ exist only in "deep" potential wells with $\log x_\mathrm{in}<-3.756$. Therefore, the standard formulation of MRI (shearing flow immersed in a constant poloidal magnetic field) does not always work in thin accretion disks. For example, in accretion disks around ordinary stars (shallow potentials), the inner disk radius may be too large for the appearance of short-wavelength unstable modes, while in accretion disks around compact stars (deep potentials), unstable modes with wavelengths smaller than the disk thickness are possible.

\subsection{Non-local modal analysis with radially changing background magnetic field.}
We have performed, for the first time, a nonlocal modal analysis of MRI with a variable background Alfvén velocity $c_A(r)$ and extensively studied the case of a power-law dependence $c_A^2(r)\propto r^{-q}$ in Keplerian flows (equations for small perturbations of the magnetic field (\ref{e:br1}) and (\ref{e:brq}), respectively). We have found that the maximum growth rate of MRI increases with increasing $q$ (Fig.~\ref{f:modesq}). Furthermore, in this case, the dimensionless potential $\tilde U_q$ \ref{e:uq} depends on both $q$ and the parameter $\epsilon=c_A^2(r_0)/v_\phi^2(r_0)$. For $q>1$, the potential $\tilde U(q,\epsilon)$ qualitatively changes with decreasing $\epsilon$ from the critical value (corresponding to the zero mode $\omega^2=0$, \Eq{e:ncAcrq}): two turning points appear to the right and left of the zero point $r_0$ (see Figures \ref{f:Uq2}-\ref{f:Uq4}). Clearly, the result of the nonlocal analysis of MRI will depend on the position of the inner boundary of the flow $x_\mathrm{in}$ relative to the zero points of the effective potential.

\begin{figure}
        \centering
        \includegraphics[width=0.7\textwidth]{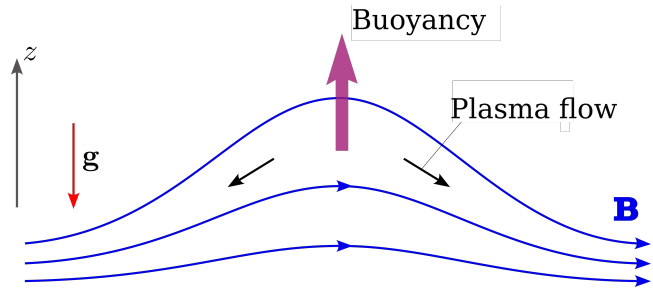}
        \caption{Schematic of the Parker instability.}
        \label{f:parker}
    \end{figure}

Thus, the nonlocal analysis of MRI demonstrates the need to consider specific features of the flow. Here, we have only considered the poloidal background magnetic field. In real situations, there may exist poloidal components as well as toroidal components of the background magnetic field. Global analysis of non-axisymmetric perturbations for MRI has been conducted, for example,  by \citet{1996MNRAS.281..119C}. The toroidal magnetic field in the disk plane should be subject to the Parker instability \citep{1966ApJ...145..811P}, which arises in the presence of gravity (see Fig.~\ref{f:parker}). A detailed description of the Parker instability in accretion disks can be found, for example, in \cite{1998Kato}. The Parker instability in accretion disks can be dubbed "magneto-gravitational instability."

\section*{Acknowledgements}
We express our gratitude to the reviewers for their careful reading of the article and their critical comments, which have significantly improved the presentation of the obtained results.
This work was supported by the Russian Science Foundation (grant number 21-12-00141). The authors would like to thank the participants of the seminars at the Department of Relativistic Astrophysics, Sternberg Astronomical Institute, Moscow State University, and the Theoretical Department of the Keldysh Institute of Applied Mathematics for valuable discussions.


\bibliographystyle{kluwer}
\bibliography{mri1}

\begin{appendix}

\section{Numerical solution of Equation~(17)
for the case \lowercase{$r_\mathrm{{out}}>{r_0}$}}

It is instructive to directly solve \Eq{e:shred} in the case when the outer boundary of the flow lies beyond the point $r_0$, where the potential $U(r_0)=0$. In dimensionless variables $x=r/r_0$ and $\tilde E=-k_z^2r_0^2$, \Eq{e:shred} reads:
\beq{b:shred}
\frac{{\rm d}^2\Psi}{{\rm d}x^2}+\left\{
\tilde E-\tilde U\right\}\Psi=0\,,
\quad \tilde U=\frac{3}{4}\frac{1}{x^2}-\frac{3}{4}\frac{1}{x^3}\, .
\eeq
The amplitude of the function $\Psi$ is arbitrary, and we choose $|\Psi(x_1)|=1$.
Conditions at the boundaries of the flow are imposed according to (\ref{e:boundary3x}).

We can simplify the numerical solution by imposing a condition not at the outer boundary point $x_\mathrm{out}>1$, but at the turning point $x_1<1$. At the turning point $x_1$, the condition $\Psi''(x_1)=0$ is automatically satisfied due to the nature of the turning point, where $\tilde U(x_1)=\tilde E$. Therefore, as a boundary condition at the turning point, we should take the value of the first derivative $\Psi'(x_1)$. It can be easily found by noticing that for small $\xi=x-x_1\ll 1$, \Eq{b:shred} reduces to the Airy equation
\beq{b:airy}
\Psi''(z)+z\Psi=0\, ,
\eeq
where 
\beq{b:z}
z=\left(\frac{3}{2}\frac{1}{x_1^3}-\frac{9}{4}\frac{1}{x_1^4}\right)^{1/3}\xi\,.
\eeq
The general solution of \Eq{b:airy} is
\beq{}
\mathrm{Ai}(z)=\frac{\sqrt{z}}{3}\left[J_{1/3}(\frac{2}{3}(-z)^{3/2})+J_{-1/3}(\frac{2}{3}(-z)^{3/2})\right]\,.
\eeq
It should be clear that if the outer boundary $x_\mathrm{out}\to\infty$, then $\Psi\to 0$, and the constant in the general solution $C_2$ must be equal to zero. 
For a finite $x_\mathrm{out}$, the constant $C_2$ is non-zero, and the value of the first derivative $\Psi'(x_1)$ should be determined from the  conditions at the outer and inner boundaries (\ref{e:boundary3x}). 
Below, we present solutions for the case of $x_\mathrm{out}\gg 1$ with the constant $C_2=0$.

At the point $x_1$, where $\xi=z=0$, the first derivative reads:
\beq{b:dpsidz}
\left.\frac{{\rm d}\Psi}{{\rm d}x}\right|_{x_1}=
\left.\frac{{\rm d}\Psi}{{\rm d}z}\frac{{\rm d}z}{{\rm d}x}\right|_{x_1}= C_1\,
\mathrm{Ai}'(0)\,\left[\frac{3}{2}\frac{1}{x_1^3}-\frac{9}{4}\frac{1}{x_1^4}\right]^{1/3}\,,
\eeq
where $\mathrm{Ai}'(0)=0.25882\ldots$ 

\begin{figure*}
\begin{center}
\includegraphics[width=0.8\textwidth]{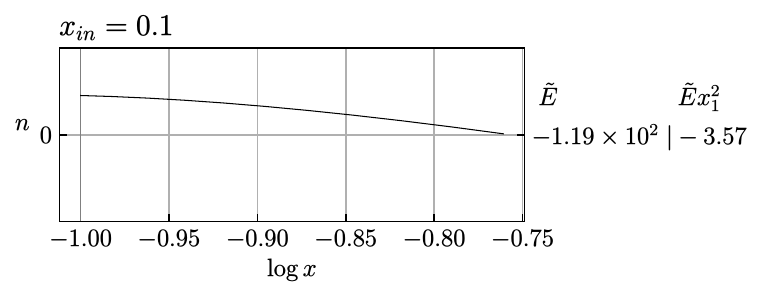}
\includegraphics[width=0.8\textwidth]{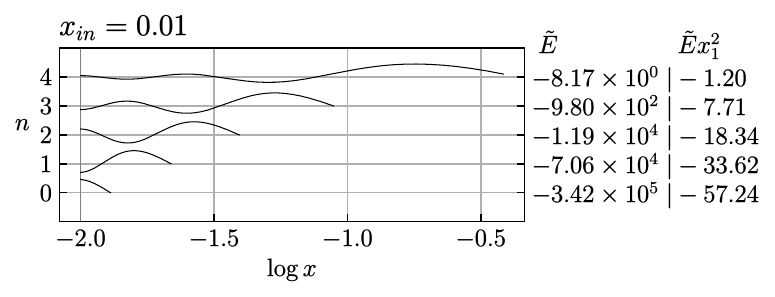}
\includegraphics[width=0.8\textwidth]{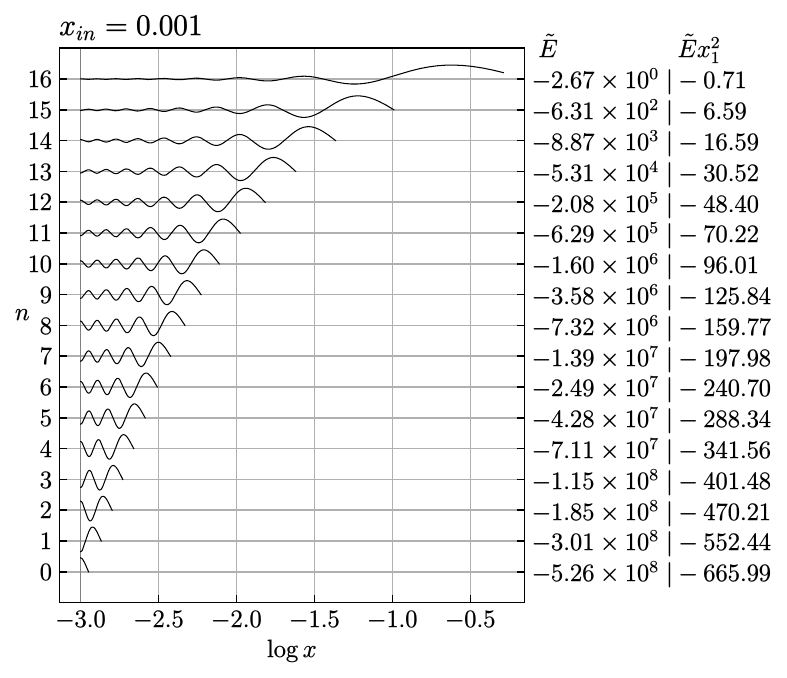}
\caption{
Normalized solutions of \Eq{b:shred}, plotted as  $\Psi/\Psi_\mathrm{max}$, with the boundary condition \eqref{e:boundary3x}, $\Psi'|_{x_\mathrm{in}}+{\Psi}/({2x_\mathrm{in}})=0$  at  $x_\mathrm{in}$, and \eqref{b:dpsidz} on $\Psi'(x_1)$  at $x_1$, corresponding to $\lim_{x\to\infty}\Psi(x)=0$, for $x_\mathrm{in}=10^{-1}$ (a), $ 10^{-2}$ (b), and $10^{-3}$ (c).
As $x_\mathrm{in}$ decreases (indicating a deeper potential well), the number of stationary negative energy levels increases. 
The exponentially decaying part of the solution for $x>x_1$ is not shown.
}
\label{f:psi_n}
\end{center}
\end{figure*}

\begin{figure*}
\includegraphics[width=0.95\textwidth]{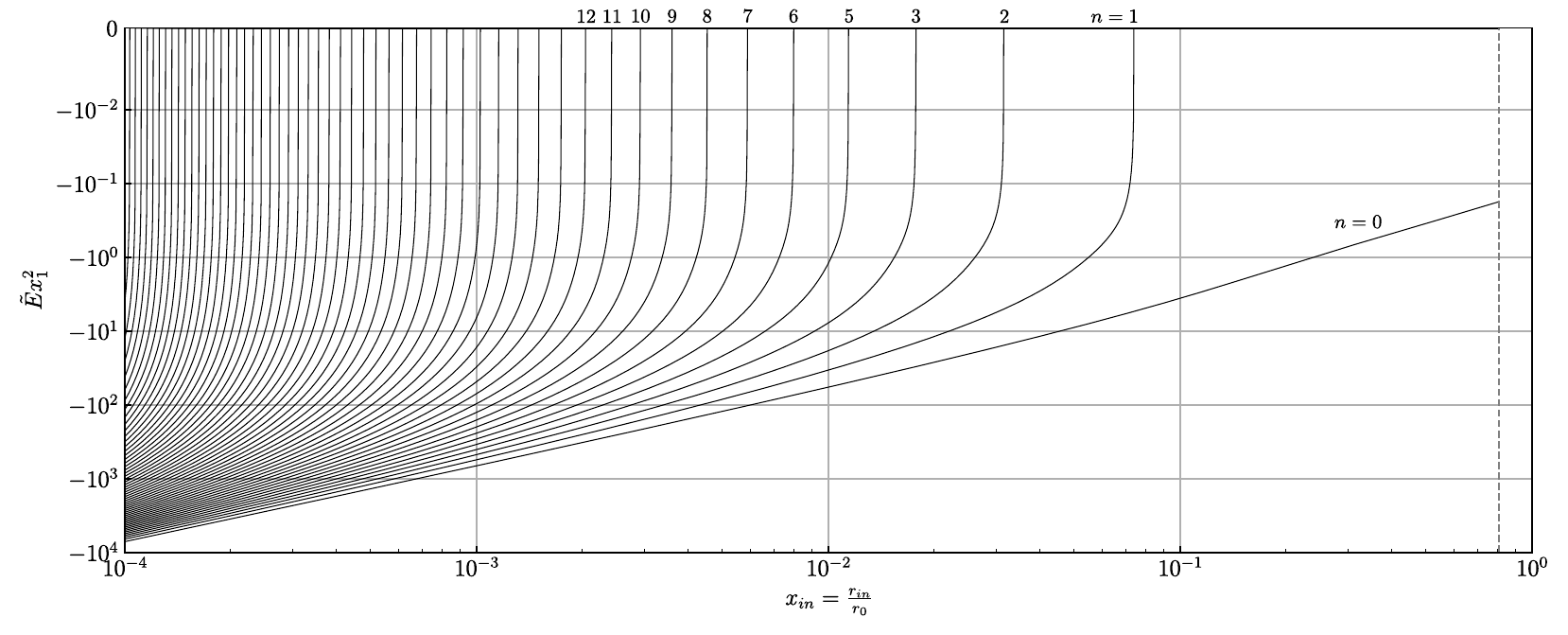}
\includegraphics[width=0.95\textwidth]{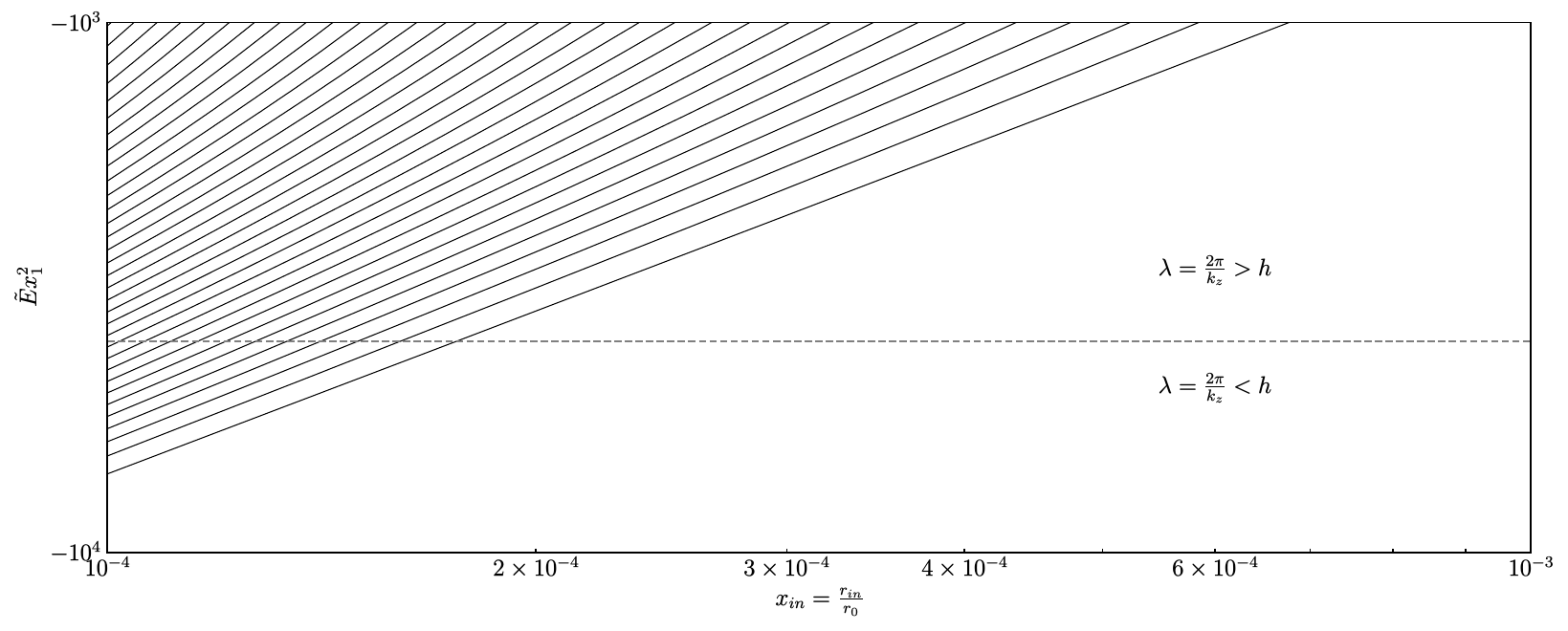}
\caption{
(a) Discrete "energy levels" $\tilde E=-(k_zr_0)^2$ of \Eq{b:shred} for different inner boundaries $x_\mathrm{in}$. As $x_\mathrm{in}$ decreases, the number of discrete "energy levels" $n=0,1,2,3\ldots$ of the function $\Psi$ increases. Deep "energy levels" with $-\tilde E x_1^2>4\times 10^3$ appear when $\log x_\mathrm{in}<-3.756$. There exists a maximum value of $x_\mathrm{in} \approx 0.8116$ with $\tilde E_0 \approx -0.2644$ for the eigenmode with $n=0$. Other modes $n=1, 2,\ldots$ can have $\tilde E=0$ at certain values of $x_\mathrm{in}$. (b) The region of "deep levels" and the boundary $\lambda=2\pi/k_z = h=0.1\, r$~ for thin accretion discs (see section \ref{s:thin_ad}).
}
\label{f:A2}
\end{figure*}

Integration of \Eq{b:shred} with the boundary conditions (\ref{e:boundary3x}) $\Psi'|_{x_\mathrm{in}}+{\Psi}/({2x_\mathrm{in}})=0$ and \Eq{b:dpsidz} for various values of $x_\mathrm{in}$ yields a family of solutions with discrete (non-equidistant) "energy levels" $\tilde E_n$ (or equivalently, according to \Eq{e:ntp}, with discrete turning points $x_1$) corresponding to the integer number $n=0,1,2,3,\ldots$ of the zeros of the function $\Psi$. 
It should be noted that for a given $x_\mathrm{in}$, there exists a different number of possible energy levels, and there is a maximum value $(x_\mathrm{in})_\mathrm{max}\approx 0.8116$ that allows the existence of a single level for $n=0$. 
The solutions, normalized to the maximum value of the function $\Psi$, are shown in Fig. \ref{f:psi_n}. The discrete "energy levels" $\tilde E=-(k_zr_0)^2$ of \Eq{b:shred} for different inner boundaries $x_\mathrm{in}$ are shown in Fig. \ref{f:A2}. The values of the energy $\tilde E$ for different levels at fixed $x_\mathrm{in}$ under various boundary conditions on the function $\Psi$ are presented in Table \ref{tab:energy_levels1}. It is evident that the type of boundary conditions has a weak influence on the energy values of the "lower" levels in deep potential wells for $x_\mathrm{in}\ll 1$.
\begin{table}
    \centering
     \caption{
     Examples of energy levels $\tilde{E}$ of \Eq{b:shred} for $x_\mathrm{in}=0.1,0.01,0.001$ and different  conditions at the inner boundary $x_\mathrm{in}$.}
     \vskip 3mm
    \label{tab:energy_levels1}
    \begin{tabular}{|r|r|r|r|}
    \hline
    $n_{\tilde{E}}$ & $\left.\left(\Psi' + {\Psi}/({2x})\right)\right\vert_{x_\mathrm{in}} = 0$ & $\Psi'|_{x_\mathrm{in}}=0$ & $\Psi|_{x_\mathrm{in}} = 0$ \\
    \hline
    \multicolumn{4}{|c|}{$x_\mathrm{in} = 0.1$} \\
    \hline
        0 & $-118.86             $ & $-86.44              $ & -- \\
        \hline
    \multicolumn{4}{|c|}{$x_\mathrm{in} = 0.01$} \\
    \hline
        4 & $-8.168              $ & $-7.096              $ & -- \\
        3 & $-9.799 \times 10^{2}$ & $-9.250 \times 10^{2}$ & $-1.229 \times 10^{2}$ \\
        2 & $-1.189 \times 10^{4}$ & $-1.145 \times 10^{4}$ & $-3.483 \times 10^{3}$ \\
        1 & $-7.057 \times 10^{4}$ & $-6.846 \times 10^{4}$ & $-2.808 \times 10^{4}$ \\
        0 & $-3.423 \times 10^{5}$ & $-3.288 \times 10^{5}$ & $-1.408 \times 10^{5}$ \\
       \hline
    \multicolumn{4}{|c|}{$x_\mathrm{in} = 0.001$}\\
    \hline
        16 & $-2.666              $ & $-2.525              $ & -- \\
        15 & $-6.314 \times 10^{2}$ & $-6.190 \times 10^{2}$ & $-7.436 \times 10^{1}$ \\
        14 & $-8.869 \times 10^{3}$ & $-8.762 \times 10^{3}$ & $-2.668 \times 10^{3}$ \\
        13 & $-5.308 \times 10^{4}$ & $-5.260 \times 10^{4}$ & $-2.266 \times 10^{4}$ \\
        12 & $-2.078 \times 10^{5}$ & $-2.063 \times 10^{5}$ & $-1.071 \times 10^{5}$ \\
        11 & $-6.287 \times 10^{5}$ & $-6.250 \times 10^{5}$ & $-3.650 \times 10^{5}$ \\
        10 & $-1.598 \times 10^{6}$ & $-1.590 \times 10^{6}$ & $-1.007 \times 10^{6}$ \\
         9 & $-3.585 \times 10^{6}$ & $-3.569 \times 10^{6}$ & $-2.397 \times 10^{6}$ \\
         8 & $-7.319 \times 10^{6}$ & $-7.290 \times 10^{6}$ & $-5.121 \times 10^{6}$ \\
         7 & $-1.390 \times 10^{7}$ & $-1.385 \times 10^{7}$ & $-1.007 \times 10^{7}$ \\
         6 & $-2.495 \times 10^{7}$ & $-2.486 \times 10^{7}$ & $-1.857 \times 10^{7}$ \\
         5 & $-4.284 \times 10^{7}$ & $-4.271 \times 10^{7}$ & $-3.258 \times 10^{7}$ \\
         4 & $-7.115 \times 10^{7}$ & $-7.094 \times 10^{7}$ & $-5.496 \times 10^{7}$ \\
         3 & $-1.155 \times 10^{8}$ & $-1.151 \times 10^{8}$ & $-9.011 \times 10^{7}$ \\
         2 & $-1.854 \times 10^{8}$ & $-1.848 \times 10^{8}$ & $-1.451 \times 10^{8}$ \\
         1 & $-3.005 \times 10^{8}$ & $-2.994 \times 10^{8}$ & $-2.328 \times 10^{8}$ \\
         0 & $-5.263 \times 10^{8}$ & $-5.227 \times 10^{8}$ & $-3.828 \times 10^{8}$ \\
        \hline
    \end{tabular}
\end{table}
\begin{table}
    \centering
     \caption{Values of the dimensionless inner flow boundary for zero energy $x_{\mathrm{in}}\vert_{\tilde{E} = 0}$ under different choices of boundary conditions. The value of $x_{\mathrm{in}}\vert_{\tilde{E} = 0}$ in the WKB approximation is given by (\ref{e:E0}) and (\ref{e:disp0}). It is noteworthy that under the third-type boundary condition (free boundary), the zero energy level is reached at a finite value of $\tilde E\approx -0.2644$ corresponding to $x_{\mathrm{in}}\approx 0.8116$. Refer to Fig.~\ref{f:A2} for further details. }
     \label{tab:energy_levels2}
     \vskip 3mm
    \begin{tabular}{r|r|r|r}
    \hline
    $n_{\tilde{E}}$ & $\left.\left(\Psi' + {\Psi}/({2x})\right)\right\vert_{x_\mathrm{in}} = 0$ & $\Psi'|_{x_\mathrm{in}}=0$ & $\Psi|_{x_\mathrm{in}} = 0$ \\
    \hline
        ... & ...                   & ...                    &  ... \\
        \hline
        16 & $1.02496 \times 10^{-3}$ & $1.02426 \times 10^{-3}$ & --- \\
        15 & $1.15546 \times 10^{-3}$ & $1.15457 \times 10^{-3}$ & $1.08531 \times 10^{-3}$ \\
        14 & $1.31260 \times 10^{-3}$ & $1.31145 \times 10^{-3}$ & $1.22775 \times 10^{-3}$ \\
        13 & $1.50417 \times 10^{-3}$ & $1.50266 \times 10^{-3}$ & $1.40021 \times 10^{-3}$ \\
        12 & $1.74105 \times 10^{-3}$ & $1.73903 \times 10^{-3}$ & $1.61176 \times 10^{-3}$ \\
        11 & $2.03875 \times 10^{-3}$ & $2.03598 \times 10^{-3}$ & $1.87519 \times 10^{-3}$ \\
        10 & $2.42010 \times 10^{-3}$ & $2.41620 \times 10^{-3}$ & $2.20897 \times 10^{-3}$ \\
         9 & $2.91976 \times 10^{-3}$ & $2.91407 \times 10^{-3}$ & $2.64061 \times 10^{-3}$ \\
         8 & $3.59244 \times 10^{-3}$ & $3.58384 \times 10^{-3}$ & $3.21255 \times 10^{-3}$ \\
         7 & $4.52880 \times 10^{-3}$ & $4.51513 \times 10^{-3}$ & $3.99300 \times 10^{-3}$ \\
         6 & $5.88807 \times 10^{-3}$ & $5.86497 \times 10^{-3}$ & $5.09744 \times 10^{-3}$ \\
         5 & $7.97202 \times 10^{-3}$ & $7.92969 \times 10^{-3}$ & $6.73423 \times 10^{-3}$ \\
         4 & $1.14131 \times 10^{-2}$ & $1.13263 \times 10^{-2}$ & $9.31161 \times 10^{-3}$ \\
         3 & $1.77392 \times 10^{-2}$ & $1.75298 \times 10^{-2}$ & $1.37231 \times 10^{-2}$ \\
         2 & $3.15576 \times 10^{-2}$ & $3.08961 \times 10^{-2}$ & $2.22593 \times 10^{-2}$ \\
         1 & $7.39693 \times 10^{-2}$ & $7.03546 \times 10^{-2}$ & $4.24509 \times 10^{-2}$ \\
         0 & $0.8116 \vert_{\tilde{E} = -0.2644}$ & $5.87626 \times 10^{-1}$ & $1.14244 \times 10^{-1}$ \\
         \hline
    \end{tabular} 
\end{table}

 It should be noted that \Eq{b:shred} can be solved using the  quasi-classical (WKB) approximation, employing the Bohr-Sommerfeld condition for discrete energy levels $E=-k_z^2<0$ with index $n=0,1,2,\ldots$ in the potential $U$ for rigid boundaries \citep{LL1974,1971pqm..book.....F}:
\beq{e:BS}
\int\limits_{r_\mathrm{in}}^{r_1}\sqrt{E-U}\,dr=
\piup\left(n+\frac{3}{4}\right)\,;\quad n=0,1,2,\ldots
\eeq
Here, $r_1$ is the turning point in the potential $U$, which can be found from \Eq{e:r1}. The existence of stationary levels with negative energy and $\omega^2<0$ indicates the instability of the flow. 
A similar equation, but with a constant slightly different from 3/4, can be obtained for a free boundary.
However, for large $n\gg 1$, all three boundary conditions ($\Psi|_{x_\mathrm{in}}=0$, $\Psi'|_{x_\mathrm{in}}=0$, $(\Psi'+{\Psi}/(2{x}))|_{x_\mathrm{in}}=0$) yield the same result.

In dimensionless units with the energy $\tilde E=-k_z^2r_0^2$ and potential $\tilde U$,  \Eq{e:BS} takes the form
\beq{e:BSx}
\int\limits_{x_\mathrm{in}}^{x_1}\sqrt{\tilde E-\frac{3}{4x^2}+\frac{3}{4x^3}}\,{\rm d}x=
\piup\,\left(n+\frac{3}{4}\right)\,.
\eeq

In the case of a large perturbation wavelength, $\tilde E=-k_z^2r_0^2\ll 1$.
At low energies $\tilde E\approx 0$, the turning point tends to the zero of the potential, $r_1\to r_0$,  and the integral in \Eq{e:BSx} is 
\beq{e:BSx0}
\int\limits_{x_\mathrm{in}}^{1}\sqrt{ -\frac{3}{4x^2}+\frac{3}{4x^3}}\,{\rm d}x=
\piup\,\left(n+\frac{3}{4}\right)\,.
\eeq
This integral is taken as follows:
\beq{e:E0}
\sqrt{3}\left[\sqrt{\frac{1}{x_\mathrm{in}}-1}+\arcsin{\sqrt{x_\mathrm{in}}}-\frac{\piup}{2}\right]=\piup\left(n+\frac{3}{4}\right)
\eeq
(compare with equation (2.24) for the case of isomoment rotation with circular velocity $v_\phi\propto 1/r$ in the original paper by \cite{Velikhov59}; here, we consider the Keplerian case with $v_\phi\propto 1/\sqrt{r}$). From \Eq{e:E0}, we find the first $x_\mathrm{in}$ for $n=0$: $x_\mathrm{in}\approx 0.575$. The exact value, which we obtained numerically (see Fig. \ref{f:A2}), is $(x_\mathrm{in})_\mathrm{max}\approx 0.8116$. Furthermore, under the given third-type boundary conditions, the energy value for the mode with $n=0$ is approximately $\tilde E[(x_\mathrm{in})_\mathrm{max}]\approx -0.2644$, and for modes $n=1,2,\ldots$, there always exists a discrete value of $x_\mathrm{in}$ for which $\tilde E=0$. For first and second-type boundary conditions, $E[(x_\mathrm{in})_\mathrm{max}]=0$ (see Table \ref{tab:energy_levels1} and \ref{tab:energy_levels2} for different conditions at the inner boundary). 
For $x_\mathrm{in}>(x_\mathrm{in})_\mathrm{max}$, the potential well becomes so shallow that there are no stationary "energy levels". It is worth noting that the accuracy of the WKB approximation increases for large $n$. In our case, for $x_\mathrm{in}=r_\mathrm{in}/r_0\ll 1$, the discrete values of $x_\mathrm{in}$ are given by:
\beq{e:disp0}
x_\mathrm{in}=
\frac{1}{\left[\left({\piup}/\sqrt{3}\right)\left(n+{3}/{4}\right)+{\piup}/{2}\right]^2+1}
\,.
\eeq
In principle, the integral in \Eq{e:BSx} can be numerically computed for any value of $\tilde E$ to find the discrete energy levels $\tilde E_n$ at fixed $x_\mathrm{in}$ (especially for large $|\tilde E|$ corresponding to thin accretion disks, as mentioned above) for any $n$. However, precise numerical solutions of \Eq{b:shred} have already been obtained.

It is worth noting that, if we disregard the term ${3}/\left({4}{x^2}\right)$ in \Eq{e:BSx}, it can be solved analytically (see \citealt{2015MNRAS.453.3257L}). 
The WKB solution is found to be in excellent agreement with the numerical results.

\section{Case \lowercase{$r_\mathrm{out}<r_0$}} 

If the outer boundary of the flow, denoted as $r_\mathrm{out}$, is smaller than $r_0$ (the point at which the potential becomes zero, $U(r_0)=0$, see \Eq{e:r0}), then the problem of finding the stationary energy levels of \Eq{b:shred} needs to be numerically solved with the boundary conditions given by (\ref{e:boundary3x}). In the case of a narrow flow (when the outer radius is close to the inner radius), the problem has an analytical solution. A similar problem, with a different rotational law and zero boundary conditions for magnetic field perturbations ($b_r=0$, corresponding to the condition $\Psi=0$), was solved by \cite{Velikhov59}. 

Let us introduce the following notations: 
\beq{}
\bar r\equiv r_\mathrm{in}+\frac{r_\mathrm{out}-r_\mathrm{in}}{2},\qquad \Delta\equiv r_\mathrm{out}-r_\mathrm{in}\ll \bar r\, .
\eeq
When the parameter $\bar r$ is fixed, \Eq{b:shred} transforms into a Sturm-Liouville problem with homogeneous boundary conditions of the third kind on the interval $x\in[a,b]$:
\begin{equation}
\label{b:SL}
\begin{cases}
&\Psi''+\lambda \Psi=0\, , \\
&\Psi'|_a+\frac{\Psi}{2a}=0\, ,  \\
&\Psi'|_b+\frac{\Psi}{2b}=0\, ,
\end{cases}
\end{equation}
where 
\beq{b:lambda}
\lambda=-k_z^2-\frac{3}{4}\frac{1}{\bar r^2}+\frac{GM}{\bar r^3}
\frac{{\omega^2}/{k_z^2}+3\,c_A^2}{\left(c_A^2-{\omega^2}/{k_z^2}\right)^2}\,.
\eeq
This problem admits the existence of only positive nontrivial eigenvalues $\lambda_n>0$, corresponding to oscillating eigenfunctions $\Psi_n$. Then, the general solution of \Eq{b:SL} is given by
\beq{}
\Psi=C_1\sin(\sqrt{\lambda}x)+C_2\cos(\sqrt{\lambda}x)\, .
\eeq
The constants $C_1$ and $C_2$ are determined by the boundary conditions. By excluding the trivial solution $\lambda=0$, we obtain the following system of equations:
\beq{b:gr}
\begin{cases}
&C_1\left(1+\frac{1}{2\sqrt{\lambda}a}\tan (\sqrt{\lambda}a)\right)+C_2\left(\frac{1}{2\sqrt{\lambda}a}-\tan (\sqrt{\lambda}a)\right)=0 \, ,\\
&C_1\left(1+\frac{1}{2\sqrt{\lambda}b}\tan (\sqrt{\lambda}b)\right)+C_2\left(\frac{1}{2\sqrt{\lambda}b}-\tan (\sqrt{\lambda}b)\right)=0\, .
\end{cases}
\eeq
System (\ref{b:gr}) has a nontrivial solution when the determinant is equal to zero. This leads to a transcendental equation for finding the eigenvalues $\lambda$:
\begin{eqnarray}
\label{b:det}
&\left[\frac{1}{2\sqrt{\lambda}b}-\frac{1}{2\sqrt{\lambda}a}\right]\,\left(1+\tan(\sqrt{\lambda}a)\,\tan(\sqrt{\lambda}b)\right)+\nonumber\\
&\left[1+\frac{1}{4\lambda ab}\right]\,
\left(
\tan(\sqrt{\lambda}a)-\tan(\sqrt{\lambda}b)
\right)=0\,.
\end{eqnarray}
Equation (\ref{b:det}) has two families of solutions:
\beq{b:2sol}
\begin{cases}
&1+\tan(\sqrt{\lambda}a)\tan(\sqrt{\lambda}b)=0\\
&-\tan(\sqrt{\lambda}(b-a))=\frac{\sqrt{\lambda}(b-a)}{1+4\lambda ab}
\end{cases}
\eeq
The first family of solutions to \Eq{b:det} is given by
\beq{b:1sol}
\cos\left(\sqrt{\lambda}\,[a-b]\right)=0\,.
\eeq
From this, we obtain the set of eigenvalues
\beq{b:lamn}
\lambda_n=\frac{\left({\pi}/{2}+n\pi\right)^2}{\Delta^2}\,,\qquad n=0,1,2,3,\ldots
\eeq
For small values of $n=1,2,3,...$, the second family of solutions (\ref{b:2sol}) can be found numerically. However, for large argument values, it quickly converges to $\lambda_n\approx \pi^2n^2/\Delta^2$ for $n\gg1$. Therefore, for large $n$, both solutions can be combined into a single expression: $\lambda_n\approx (\pi^2/4)(1+n)^2/\Delta^2$.
It should be noted that under first-kind homogeneous boundary conditions $\Psi|_{a,b}=0$ or second-kind homogeneous boundary conditions $\Psi'|_{a,b}=0$, the following solutions are obtained:
\begin{eqnarray}
&\sin\sqrt{\lambda}(b-a)=0\quad\to\quad \lambda_n=\frac{\pi^2n^2}{\Delta^2}\, , \nonumber\\
&\sin\sqrt{\lambda}(a+b)=0\quad\to\quad\lambda_n=\frac{\pi^2n^2}{(a+b)^2}\, ,  \nonumber\\
&n=1,2,3,\ldots
\end{eqnarray}

From (\ref{b:lamn}), we obtain a set of equations for each eigenvalue $\lambda_n$:
\begin{eqnarray}
&-k_z^2-\frac{3}{4}\frac{1}{\bar r^2}+\frac{GM}{\bar r^3}
\frac{{\omega^2}/{k_z^2}+3c_A^2}{\left(c_A^2-{\omega^2}/{k_z^2}\right)^2}= \lambda_n
\approx\frac{\pi^2(n+1)^2}{4\Delta^2}
\,,\nonumber\\  &n=0,1,2,3,\ldots
\end{eqnarray}
Hence, we obtain the dispersion equations in the following form:
\begin{eqnarray}
\label{b:dispur}
    &\frac{\omega^2}{c_A^2\,k_z^2}=
\left(1+\frac{K}{2}\right)\,\left[1\pm\sqrt{1+\frac{4(3K-1)}{(K+2)^2}}\,\right]\, ,\nonumber\\
&K\approx\frac{v_\phi^2(\bar r)}{c_a^2}\frac{1}{{\pi^2}/{4}\,(n+1)^2\left({\bar r}/{\Delta}\right)^2+k_z^2\bar r^2+{3}/{4}}\, .
\end{eqnarray}
For unstable modes with $\omega^2<0$, in the dispersion equation (\ref{b:dispur}), we keep the minus sign before the square root and require $K>1/3$, which leads to the appearance of a critical magnetic field for the MRI mode:
\beq{b:cacr3}
c_A^2<(c_A^2)_{cr}\approx \frac{4v_\phi^2(\bar r)}{({1}/{3})\pi^2(n+1)^2\left({\bar r}/{\Delta}\right)^2+\frac{4}{3}k_z^2\bar r^2+1}\,.
\eeq
It should be noted that for the first and second kind boundary conditions, the critical field values are given by
\beq{}
\begin{cases}
    &(c_A^2)_{cr}=\frac{4v_\phi^2(\bar r)}{({4}/{3})\pi^2n^2\left({\bar r}/{\Delta}\right)^2+({4}/{3})k_z^2\bar r^2+1}\,,\quad \Psi|_{a,b}=0\\
    &(c_A^2)_{cr}=\frac{4v_\phi^2(\bar r)}{({1}{3})\pi^2n^2+({4}/{3})k_z^2\bar r^2+1}\,,\quad \Psi'|_{a,b}=0
\end{cases}
\eeq
The similarity of results between the first and third kind boundary conditions is notable.

Lastly, it is noteworthy that for the case of \hbox{$r_\mathrm{out}<r_0$} (when the potential has the same sign over the entire interval $[r_\mathrm{in}, r_\mathrm{out}]$), the asymptotic behavior of eigenvalues is given by $\lambda_n\sim \pi^2n^2/(r_\mathrm{out}-r_\mathrm{in})^2$ for large $n\gg 1$, which follows from the general theorems of the Sturm-Liouville problem for any continuous potential with a consistent sign (see, for example, \citealt{2012sturm}).

\end{appendix}
\label{lastpage}
\end{document}